\documentclass[onecolumn,10pt,a4paper,sort&compress,aps,pre,showpacs,superscriptaddress]{revtex4-2}

\usepackage[english]{babel}

\usepackage[caption=false]{subfig}
\addto\captionsenglish{}

\usepackage{overpic}
\usepackage{amsmath,amssymb,amsthm}
\usepackage[italicdiff]{physics}
\usepackage{commath}
\usepackage{graphicx}% Include figure files
\usepackage{dcolumn}% Align table columns on decimal point
\usepackage{bm}% bold math
\usepackage{hyperref}% add hypertext capabilities
\usepackage{etoolbox} % for \appto
\usepackage{siunitx}
\usepackage{multirow}
\usepackage[capitalise,nameinlink]{cleveref}
\newcommand{\aref}[1]{\hyperref[#1]{Appendix}}
\usepackage{blkarray}

\usepackage[normalem]{ulem}
\usepackage{comment}
\usepackage[]{xcolor}
\newcommand \bl{\color{blue}}

\newcommand \rd{\color{red}}

\newcommand \Pe {\textrm{Pe}}

\begin{document}

%%%% Article title to be placed here
\title{Generalised Taylor dispersion of chiral microswimmers}

\author{Keito Ogawa}
%\email{ishimoto@kurims.kyoto-u.ac.jp}
%\affiliation{Research Institute for Mathematical Sciences, Kyoto University, Kyoto 606-8502, Japan}

\author{Kenta Ishimoto}
\email{ishimoto@kurims.kyoto-u.ac.jp}
%\affiliation{Research Institute for Mathematical Sciences, Kyoto University, Kyoto 606-8502, Japan}

%%%%%%%%% Insert author address here
\address{Research Institute for Mathematical Sciences, Kyoto University, Kyoto 606-8502, Japan}

%%%% Abstract text to be placed here %%%%%%%%%%%%
\begin{abstract}
Transport phenomena of microswimmers in fluid flows play a crucial role in various biological processes, including bioconvection and cell sorting. In this paper, we investigate the dispersion behavior of chiral microswimmers in a simple shear flow utilizing the generalized Taylor dispersion (GTD) theory, motivated by biased locomotion of bacterial swimmers known as bacterial rheotaxis. We thus focus on the influence of shear-induced torque effects due to particle chirality, employing an extended Jeffery equation for individual deterministic dynamics. We then numerically calculate macroscopic parameters including averaged swimming velocity and effective diffusion tensor using spherical harmonic expansion, and argue the obtained results based on the fixed points and their stability of the orientational dynamical systems. Our results reveal that chiral effects induce biased locomotion and we observe qualitative transitions in the orientational distribution with increasing Pecl\'et number, aligning with previous experimental findings. The diffusion tensor analysis highlights significant reduction in the diffusion coefficient perpendicular to the flow plane due to chirality. This suggests potential applications in flow-mediated cell separation and we numerically demonstrate such chirality-induced fluid transportation. The presented methods will be useful in predicting and controlling dispersion behaviors of such chiral microswimmers.
\end{abstract}
%%%%%%%%%%%%%%%%%%%%%%%%%%%

\maketitle

\section{Introduction}

Transport of particles in fluid flow has been an important research subject in biological fluid dynamics for understanding and controlling mixing, diffusion and separation processes of colloidal particles and suspensions, including cancer and red blood cells \cite{Koumoutsakos2013, Fogelson2015, Secomb2017, Takeishi2019}. These two decades attract research interests in the transportation of a self-propelled particle and its suspensions  \cite{Lauga2020, Bees2020, Ishimoto2023}, being motivated by high-throughput image data of swimming bacteria and eukaryotic planktons as well as precise manipulation of these cells by microfluidic devices.  

In the fluid transport of these microswimmers, their orientational dynamics become more significant in addition to the dynamics of particle position, since it determines the swimming direction and directly affects the particle transport. Many biological swimmers in the fluid flow then exhibit various trajectory patterns as a result of their orientational dynamics and active motions. Indeed, in the vicinity of a wall boundary, some swimming cells have been known to swim upstream under a flow, which is often referred to as rheotaxis. Examples include bacteria \cite{Kaya2012, Mathijssen2019}), sperm cells \cite{Kantsler2014, Ishimoto2015, Phuyal2022} and artificial Janus particles \cite{Katuri2018, Sharan2022}, and this upstream migration are known to be a result of fluid-particle-wall interactions.

In a bulk shear flow, on the other hand, bacterial cells exhibit perpendicular migration to the flow plane, which is called a bacterial bulk rheotaxis \cite{Marcos2012, Jing2020, Zhang2021, Ronteix2022}. A spheroidal particle in an inertialess regime periodically rotates in a simple shear flow and its orientational vector forms a closed loop, which is well-known as the Jeffery orbits \cite{Jeffery1922}. This periodic rotation is described by an exact solution to the Stokes equation of the low-Reynolds-number flow. Recent studies found that the key physics behind the bacterial bulk rheotaxis is the chirality of the particle, which is not considered in the Jeffery orbits. Indeed, swimming bacterial cells possess one or multiple helical appendages called flagella and such a chiral object experiences a fluid torque, which turns the cells away from the shear plane, in other words, to be aligned parallel to the background vorticity vector. 
This chirality-induced torque is recently theoretically formulated through hydrodynamic shape theory, which is based on the invariant properties of resistance tensors under a particular change of coordinates \cite{Brenner1964b, Brenner1964c}. Ishimoto \cite{Ishimoto2020, Ishimoto2020b} considered an arbitrary axisymmetric hydrodynamic shape and extended the Jeffery equation into chiral particles, which include a simple helix and bacterial cells. 

\begin{figure}[t]
    \centering
    \includegraphics[keepaspectratio, scale=0.4]{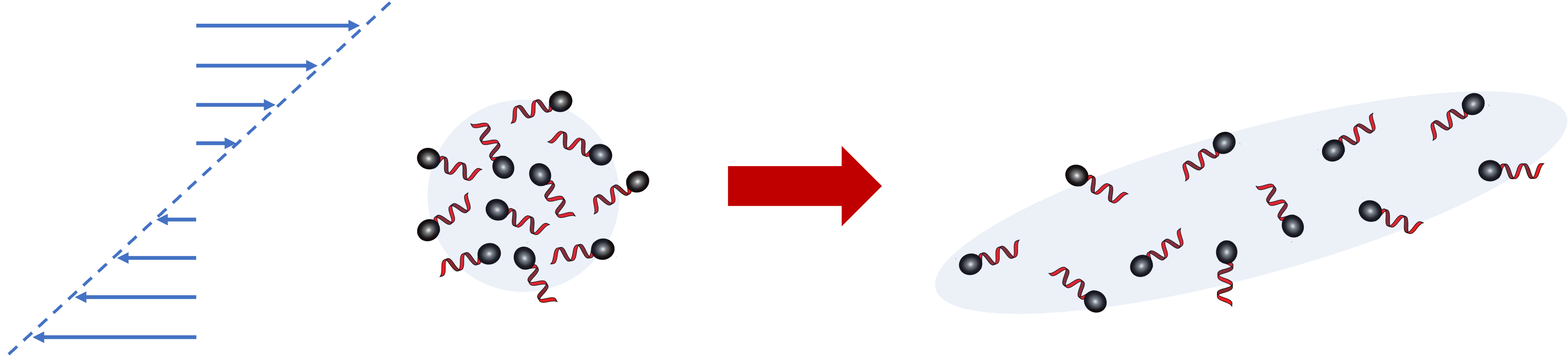}
    \caption{Schematic of Taylor dispersion of chiral microswimmers by a shearing flow. }
    \label{taylor-dispersion}
\end{figure}

As illustrated in Fig.\ref{taylor-dispersion}, these chiral microswimmers are transported in a fluid flow and, at the same time, the swimmer disperse due to the thermal and active orientational diffusion. 
To analyse these stochastic processes, we consider a probability distribution function in physical and orientation phase space, which is in general 6-dimensional, offering a numerical challenge \cite{WJC2022}. Here, focusing on a macroscopic diffusion process of chiral microswimmers, we study the generalised Taylor dispersion (GTD) theory \cite{FB1991,FB1993}, which extends the classical Taylor-Aris theory \cite{Taylor1953, Aris1956} to include internal degrees of freedom such as orientation, to bypass the large-dimensional computations.

The GTD theory has been applied to analyse population-level dispersion behaviours of swimming microorganisms in shear flows, in particular in the context of bioconvention \cite{Bees2020}. Hill \& Bees \cite{HB2002} considered a spherical microswimmer that exhibits a gyrotactic response to swim upwards in a simple shear under gravity. Bearon \cite{Bearon2003} analysed a macroscopic diffusion of spherical microswimmers with chemotactic responses. Manela \& Frankel\cite{MF2003} comprehensively analysed non-spherical microswimmers with gyrotactic response to estimate the macroscopic diffusion tensor, pointing out numerical difficulty at high Pecl\'et number regime, where the shear strength dominates the individual swimmer diffusion. The GDT theory has also been used for microswimmers exhibiting chemotaxis \cite{Bearon2003, BC2012, TL2018}, emphasising the wide applicability of the approach.

In this paper, motivated by bacterial bulk rheotaxis, we will examine the population-level dispersion phenomena of chiral microswimmers in a shear flow, by using the GTD theory to quantitatively estimate macroscopic transport velocity of the distribution and its diffusion coefficients. Using the extended Jeffery equation for particle dynamics in simple shear, we will provide a general theoretical framework and present numerical demonstrations that quantify the dispersive transport of chiral swimmers.

The contents of the paper are as follows. 
In Sec. \ref{sec:prob}, we introduce the individual deterministic dynamics of a general chiral microswimmer in a shear flow and the GTD theory in our problem. We also present a summary of numerical methods for quantifying the macroscopic model parameters such as averaged swimming velocity and efficient diffusion tensor. We present our numerical results on the orientational distribution function in Sec. \ref{sec:ori}, after theoretically analysing the structure of the deterministic dynamical systems and the stability of the fixed points. In Sec. \ref{sec:disp}, we discuss the dispersion behaviours of the chiral microswimmers with the numerical results of effective diffusion tensor and dispersion in physical coordinates. Conclusions are made in Sec. \ref{sec:conc}. 

\section{Problem setup}
\label{sec:prob}

\subsection{Chiral active particles in a simple shear}
%\begin{itemize}
%    \item Helicoidal symmetry (chirality)
%    \item Linear and angular motions of (active) chiral particles
%    \item Some remarks on chiral parameters
%\end{itemize}

\begin{figure}[t]
\centering
\begin{overpic}[width=5cm]{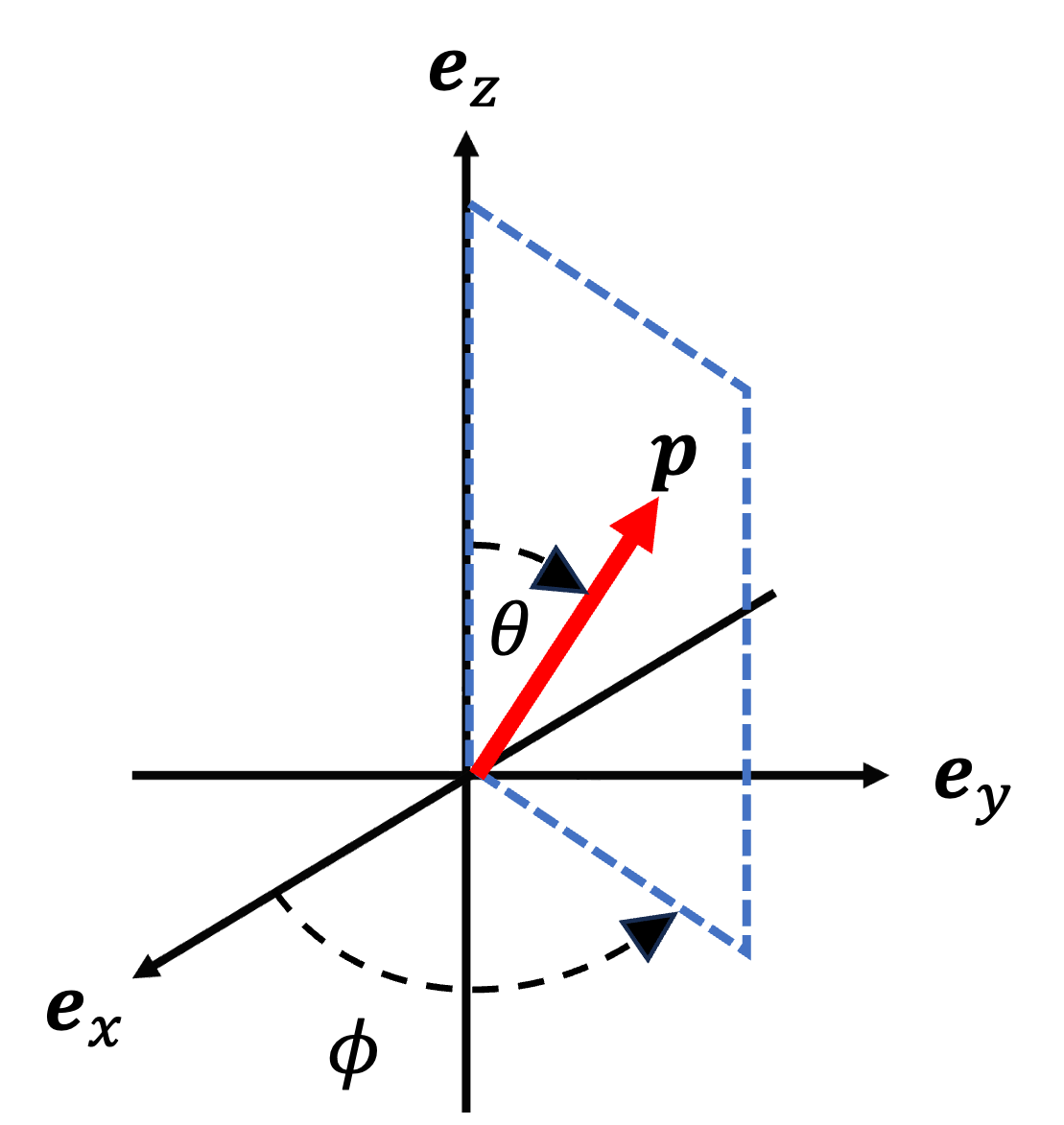}
\put(10,76){(a)}
\end{overpic}
\begin{overpic}[width=4cm]{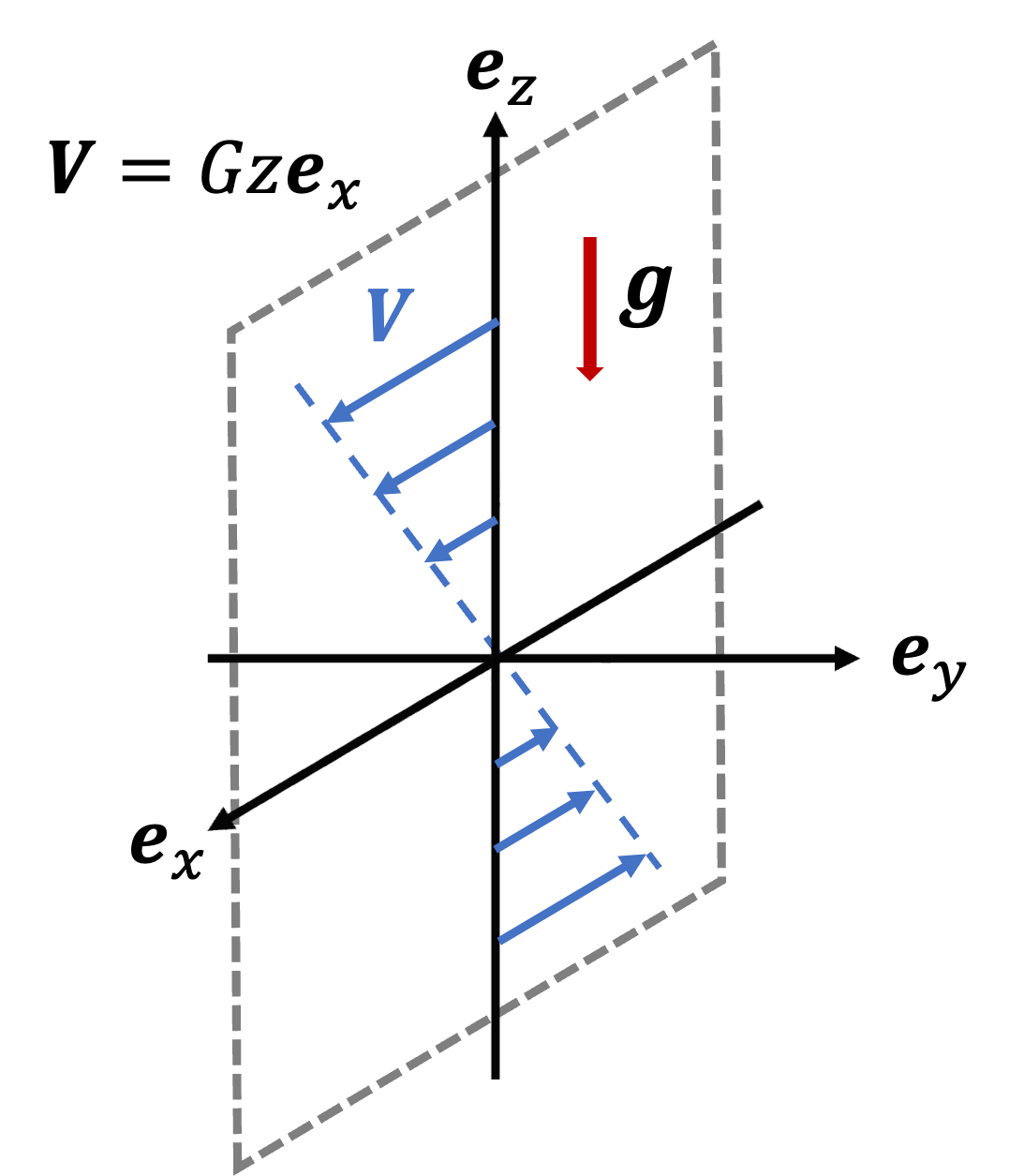}
\put(-5,85){(b)}
\end{overpic}
\caption{Schematic of coordinates and shear flows. (a) The polar coordinates for the particle orientation vector $\bm{p}$. (b) External shear flow is set as $\bm{V}=Gz\bm{e}_{x}$ with its shear strength $G$, and the gravity $\bm{g}$ is taken downwards (negative $\bm{e}_{z}$ axis).}
\label{fig_micro}
\end{figure}

We consider a microswimmer that is moving along its orientation vector $\bm{p}$ with its speed $V_s$ in a quiescent flow. As in Fig. \ref{fig_micro}(a), we introduce the laboratory frame $\{\bm{e}_x, \bm{e}_y, \bm{e}_z\}$ and polar coordinates $(\theta, \phi)$ to describe the swimmer orientation vector $\bm{p}$. We assume that the particle is placed in a linear background flow $\bm{V}(\bm{x})$ with $\bm{x}=(x, y, z)^{\textrm{T}}$. The background vorticity vector $\omega_i=\epsilon_{ijk}(\partial V_k/\partial x_j)$ ($i,j,k\in \{x,y,z\}$) and the rate-of-strain $E_{ij}=
(\partial V_i/\partial x_j+\partial V_j/\partial x_i)/2$ are constants over a space, where $\epsilon_{ijk}$ is the Levi-Civita symbol and the Einstein summation convention is used throughout this paper.

We assume that the swimmer is neutrally buoyant but that the centre of geometry does not always coincides with the centre of gravity as usually assumed for gyrotactic swimming microorganisms. As in Fig. \ref{fig_micro}(b), we take the gravitational acceleration towards $-\bm{e}_z$ axis, and the background flow  as a simple linear shear via $\bm{V}=G z\bm{e}_x$. This flow configuration has been studied in the context of swimming algae in the upper ocean and is referred to as a vertical shear \cite{HB2002}.

The angular dynamics of a microscopic spheroidal object in a linear background flow follows the well-known Jeffery equation \cite{Jeffery1922}. Chiral objects such as a helix experience shear-induced hydrodynamic forces and torques, and recently the Jeffery equation was extended to such chiral objects, given by \cite{Ishimoto2020}
\begin{equation}
    \dot{\bm{p}}=\frac{1}{2B}\left[\bm{e}_{z}-(\bm{e}_{z}\cdot\bm{p})\bm{p}\right]+\frac{1}{2}\bm{\omega}\times\bm{p}+b(\mathsf{I}-\bm{p}\bm{p})\cdot\mathsf{E}\cdot\bm{p}+c\left[(\mathsf{I}-\bm{p}\bm{p})\cdot\mathsf{E}\cdot\bm{p}\right]\times\bm{p} \label{Jeffery},
\end{equation}
where the dot symbol of the left-hand side denotes the time derivative and $\mathsf{I}$ is the identity tensor. The first term in the right-hand side of Eq. \eqref{Jeffery} represents the alignment towards the $\bm{e}_z$ axis by the gravitational torque and $B$ is the time-scale for this alignment. The second term is the rotation by the local background vorticity. The third and fourth terms include the shape effects. The parameter $b$ represents the effective aspect ratio and is known as the Bretherton parameter \cite{Bretherton1962}. This value typically ranges from -1 to 1, with $b=0$ for a sphere, $b\rightarrow 1$ at the slender limit and  $b\rightarrow -1$ for the disk limit. The parameter $c$ represents the strength of the chirality-induced rotation and thus vanishes for an achiral particle such as a body of revolution.

Particle chirality also induces shear-induced drift velocity $\bm{U}_s(\bm{p})$ and the total velocity is the sum of the background flow as $\bm{U}=Gz\bm{e}_x+\bm{U}_s$. Its general form is explicitly given by \cite{Ishimoto2020, Ishimoto2020b}, with extra terms containing shape parameters as in Eq. \eqref{Jeffery}. %{\rd [I have tentatively removed the parts on the chirality-induced drift from the paper to avoid unnecessary confusions for readers. If we could compute the effects of the chirality-induced drift, I'll bring the removed parts into the main text.]} 
The details of each shape parameter associated with the shape symmetry are summarised in the Appendix of Dalwadi et al. \cite{Dalwadi2024b}, where parameter estimates are also provided for a model bacterium cell. For a typical bacterial swimmer, however, these additional drift terms are significantly small compared with the swimming velocity. Hence, we neglect these additional terms as in the previous study \cite{Jing2020}, and we simply set $\bm{U}_s=V_s\bm{p}$ with $V_s$ being the swimming velocity.

%\begin{equation}   \bm{U}=Gz\bm{e}_x+V_s\bm{p}+\gamma(\mathsf{I}-\bm{p}\bm{p})\cdot\mathsf{E}\cdot\bm{p}-\beta\left[(\mathsf{I}-\bm{p}\bm{p})\cdot\mathsf{E}\cdot\bm{p}\right]\times\bm{p}-\delta (\bm{p}\cdot\mathsf{E}\cdot\bm{p}) \bm{p}\label{Jeffery2}.
%\end{equation}
%The first and second terms of the right-hand side of Eq.\eqref{Jeffery2} are the drift by the background flow and the swimming velocity, respectively. The remaining three terms are due to the shape of the particle. $\beta$ is a shape parameter corresponding to chiral effects and this vanishes for an achiral particle. $\gamma$ and $\delta$ are shape parameters corresponding to fore–aft asymmetry effects and thus vanish for a particle with fore–aft symmetry. The details of each shape parameter associated with the shape symmetry are summarised in the Appendix of Dalwadi et al. \cite{Dalwadi2024b}, where parameter estimates for a model bacterium cell are also provided.

We introduce the probability distribution function for the swimmer whose position and orientation are $\bm{R}'$ 
 and $\bm{p}'$ at time $t=0$ and  $\bm{R}$ 
 and $\bm{p}$ at time $t$ by $P(\bm{R},\bm{p},t|\bm{R}^{\prime},\bm{p}^{\prime})$. The Fokker-Planck equations for this distribution therefore satisfy
\begin{eqnarray}
    \frac{\partial P}{\partial t }+\nabla_{\bm{R}}\cdot\bm{J}+\nabla_{\bm{p}}\cdot\bm{j}=0, \label{P}
\end{eqnarray}
with $\nabla_{\bm{R}}$ and $\nabla_{\bm{p}}$ denoting the gradients in space and orientation, defined as $\nabla_{\bm{R}}=\bm{e}_{x}\frac{\partial}{\partial x}+\bm{e}_{y}\frac{\partial}{\partial y}+\bm{e}_{z}\frac{\partial}{\partial z}$ and $\nabla_{\bm{p}}=\bm{e}_{\theta}\frac{\partial}{\partial \theta}+\bm{e}_{\phi}\csc\theta\frac{\partial}{\partial\phi}$, respectively. The probabilistic current in space and orientation are obtained as
\begin{equation}
        \bm{J}=\bm{U}(\bm{R}, \bm{p})P ~\textrm{and}~~
    \bm{j}=\dot{\bm{p}}(\bm{p})P-d_{r}\nabla_{\bm{p}}P \label{Jj},
\end{equation}
where $d_r$ is the rotational diffusion constant and we neglected the translational diffusion for brevity as this effect is less important in many swimming microorganisms compared with the rotational diffusion \cite{HB2002}.

We now introduce a macroscopic description of the collective dynamics of the suspensions of the chiral microswimmers, by considering the position distribution function, which is obtained by integrating over the orientational space $\mathbb{S}^2$ as
\begin{align}
    \bar{P}(\bm{R},t|\bm{R}^{\prime},\bm{p}^{\prime})=\int_{\mathbb{S}^2}P(\bm{R},\bm{p},t|\bm{R}^{\prime},\bm{p}^{\prime})d^2\bm{p}.
    \label{PB}
\end{align}
We now assume that the position distribution $\bar{P}$ follows the Focker-Planck equation,
\begin{equation}
    \pdv{\bar{P}}{t}+\nabla_{\bm{R}}\cdot\bm{\bar{J}}=0, \label{FPB}
\end{equation}
with the orientation-averaged probabilistic current
\begin{equation}
    \bar{\bm{J}}=\left[\bm{V}(\bm{R})+\bar{\bm{U}}\right]\bar{P}-\bar{\mathsf{D}}\cdot\nabla_{\bm{R}}\bar{P}. \label{JB}
\end{equation}
As the initial condition for $\bar{P}$, we consider a point distribution and assume that the distribution function exponentially decays at the far field.
%As the initial condition for $\bar{P}$, we consider a point distribution
%\begin{equation}
%    \bar{P}(t=0)=\delta(\bm{R}-\bm{R}^{\prime})
%\end{equation}
%and we assume that the distribution function  exponentially decays at the far field to guarantee 
%\begin{equation}
%    \bar{P}|\bm{R}-\bm{R}^{\prime}|^{m}\rightarrow 0~\textrm{and}~~ \bar{\bm{J}}|\bm{R}-\bm{R}^{\prime}|^{m}\rightarrow\bm{0}~~\textrm{as} ~~|\bm{R}-\bm{R}^{\prime}|^{m}\rightarrow\infty.
%\end{equation}
%for an arbitrary positive integer $m$.

The collective dynamics of the microswimmers are therefore described by a convection-diffusion equation for the averaged swimmer density $N(\bm{R},t)$ with averaged swimming velocity $\bar{\bm{U}}$ and effective diffusion tensor $\bar{\mathsf{D}}$, via
\begin{equation}
    \pdv{N}{t}=-\nabla\bigg[N(\bm{V}+\bar{\bm{U}})-\bar{\mathsf{D}}\cdot\nabla N\bigg]
\end{equation}

To obtain the macroscopic model parameters $\bar{\bm{U}}$ and $\bar{\mathsf{D}}$ from individual swimmer dynamics Eqs. \eqref{P}-\eqref{Jj}, we employ the generalised Taylor dispersion (GTD) theory developed by Frankel \& Brenner \cite{FB1989, FB1991}.

\subsection{Generalised Taylor dispersion theory}
%\begin{itemize}
    %\item Fokker-Planck equation (with respect to $P(\bm{R}, \bm{p}, t)$ and $\overline{P}(\bm{R},t)$)
    %\item $P_0^{\infty}$, $\overline{U}$, $\overline{\mathsf{D}}$
    %\item Remarks on the additional term to guarantee the positive-definiteness of the diffusion tensor.
%\end{itemize}

We assume that the probability distribution function $\bar{P}(\bm{R},t|\bm{R}^{\prime},\bm{p}^{\prime})$ is well approximated by its long-time asymptotic, i.e., $ d_{r}t\gg 1$. The GTD theory then provides a closed form of equations that compute the averaged swimming velocity and the effective diffusion tensor. Since the background flow field is uniform in space, the probability distribution $\bar{P}(\bm{R},t|\bm{R}^{\prime},\bm{p}^{\prime})$ may be a function $\tilde{P}(\bm{R},t)$ that is independent of the initial distribution $\bm{R}^{\prime},\bm{p}^{\prime}$. Indeed, when the background flow $\bm{V}$ is a simple linear shear, the velocity fields are written as $\bm{V}(\bm{R})=\bm{V}(\bm{R}^{\prime})+(\bm{R}-\bm{R}^{\prime})\cdot\vb{G}$ and we introduce a change of coordinates via
\begin{align}
    \bm{R}^{(1)}=(\bm{R}-\bm{R}^{\prime})\cdot e^{-\vb{G}t}-[\bm{V}(\bm{R}^{\prime})+\bar{\bm{U}}]\cdot\int_{0}^{t}e^{-\vb{G}t_{1}}dt_{1}
    \label{R1},
\end{align}
to rewrite the Fokker-Planck equation for $\bar{P}$, \eqref{FPB}-\eqref{JB}. The probabilistic current \eqref{JB} then becomes
\begin{align}
    \bar{\bm{J}}(\bm{R}^{(1)})=-e^{-\mathsf{G}^{\text{T}}t}\cdot\bar{\mathsf{D}}\cdot e^{-\mathsf{G}t}\cdot\nabla_{\bm{R}^{(1)}}\bar{P},
\end{align}
without the drift terms. Hereafter, however, we omit the upper suffix $^{(1)}$ throughout the paper, following the standard usage \cite{FB1989, FB1991}.

With writing the $m$-th moment of the distribution $P$ as a $m$-rank tensor,
\begin{align}
    \mathsf{M}_{m}=\int_{\mathbb{R}^3}\int_{\mathbb{S}^2}(\bm{R}-\bm{R}^{\prime})^m Pd^2\bm{p}d^3\bm{R}
\end{align}
we may define the averaged velocity and the effective diffusion tensor via the first and second moments as
\begin{equation}    \bar{\bm{U}}=\lim_{t\rightarrow\infty}\frac{\delta \bm{M}_{1}}{\delta t},
    %\label{defU}
~\textrm{and}~~\bar{\mathsf{D}}=\lim_{t\rightarrow\infty}\frac{1}{2}\frac{\delta}{\delta t}(\mathsf{M}_{2}-\bm{M}_{1}\bm{M}_{1})
    \label{defD},
\end{equation}
respectively. Here, the derivative represents the Oldroyd derivative in the co-deforming frame with the fluids, given by 
$\delta\mathsf{A}/\delta t=D\mathsf{A}/Dt-\mathsf{G}\cdot\mathsf{A}-\mathsf{A}\cdot\mathsf{G}^{\text{T}}$ for an arbitrary tensor $\mathsf{A}$, where $D/Dt$ represents the Lagrangian (material) derivative and $\mathsf{G}$ is the velocity gradient tensor whose components are given by $G_{ij}=\partial V_{j}/\partial x_{i}$.

The spatial average of the probability distribution \begin{align}  P_{0}=\int_{\mathbb{R}^3}P(\bm{R},\bm{p},t|\bm{R}^{\prime},\bm{p}^{\prime})d^3\bm{R}
    \label{P0def}
\end{align}
has its long-time asymptotic form,
\begin{align}
    P_{0}^{\infty}(\bm{p})=\lim_{t\rightarrow\infty}\int_{\mathbb{R}^3}P(\bm{R},\bm{p},t|\bm{R}^{\prime},\bm{p}^{\prime})d^3\bm{R}
    \label{P0inf}
\end{align}
which satisfies, according to the GTD theory \cite{FB1989, FB1991},
\begin{equation}
    \nabla_{\bm{p}}\cdot(\dot{\bm{p}}P_{0}^{\infty}-d_{r}\nabla_{\bm{p}}P_{0}^{\infty})=0
    \label{eqP0},
\end{equation}
with the normalisation condition,
\begin{equation}
\int_{\mathbb{S}^2}P_{0}^{\infty}d^2\bm{p}=1
\label{P0norm}.  
\end{equation}
The averaged velocity and the effective diffusion constant are provided as 
\begin{equation}
    \bar{\bm{U}}=\int_{\mathbb{S}^2}P_{0}^{\infty}(\bm{p})\bm{U}_s(\bm{p}) d^2\bm{p},
    %\label{UB}\\
    ~\textrm{and}~~
        \bar{\mathsf{D}}=\int_{\mathbb{S}^2}[V_{s}P_{0}^{\infty}\bm{B}\bm{p}+P_{0}^{\infty}\bm{B}\bm{B}\cdot\vb{G}]^{\text{sym}}d^2\bm{p}
    \label{nDB},
\end{equation}
where `$\textrm{sym}$' indicates its symmetric part and the vector field $\bm{B}(\bm{p})$ is introduced as the long-time limit of the spatial deviation of a particle with its orientation $\bm{p}$ from its mean, given by
\begin{equation}
    \bm{B}(\bm{p})=\lim_{t\rightarrow\infty}\left(\frac{\bar{\bm{P}}_{1}}{P_{0}}-\bm{M}_{1}\right)
~\textrm{with}~~
    \bar{\bm{P}}_{1}(\bm{p},t|\bm{p}^{\prime})=\int_{\mathbb{R}^3}(\bm{R}-\bm{R}^{\prime})Pd^3\bm{R}.
\end{equation}
The vector field $\bm{B}(\bm{p})$ then satisfies \cite{FB1991}
\begin{align}
    &\nabla_{\bm{p}}\cdot[\dot{\bm{p}}P_{0}^{\infty}\bm{B}-d_{r}\grad_{p}(P_{0}^{\infty}\bm{B})]-P_{0}^{\infty}\bm{B}\cdot\vb{G}=P_{0}^{\infty}(\bm{U}_s-\bar{\bm{U}})
    \label{nB},
\end{align}
with a constraint 
\begin{align}
&\int_{\mathbb{S}^2}P_{0}^{\infty}\bm{B}d^2\bm{p}=0
    \label{Bn},
\end{align}
from the normalisation of $P_{0}^{\infty}$.
The second term in \eqref{nDB} is the correction due to the background shear and the associated coordinate transformation \eqref{R1}, and  guarantees the positive-definiteness of the diffusion tensor \cite{HB2002}.

We will first solve \eqref{eqP0} and the associated normalisation condition to obtain $P_0^\infty$, which allows us to compute $\bar{\bm{U}}$ in \eqref{nDB}. Then by substituting $P_0^\infty$ into \eqref{nB}, we will solve $\bm{B}$ together with the condition \eqref{Bn} to obtain $\bar{\mathsf{D}}$ in \eqref{nDB}. The detailed methods for these computations will be discussed in the next section.

\subsection{Numerical methods}
%\begin{itemize}
%    \item Coordinates setting
%    \item Non-dimensinalisation
%    \item Parameter values for the later sections
%    \item Spherical harmonics expansions
%    \item Linear equations
%\end{itemize}

To non-dimensionalise the system, %we first introduce the shear strength  $G=\norm{\mathsf{G}}=\sqrt{G_{ij}G_{ij}}$, and
we use the shear strength as the unit of time as $T=G^{-1}$. We employ the unit of length as $L=V_s/d_r$, which naturally introduces the non-dimensional parameter of the system, known as the Pecl\'et number $\text{Pe}=G/d_{r}$. %This formulation may not be compatible with non-motile particles. In that case, we use particle size for the unit of length. Note, however, that we only present the results for non-motile particles with physical units.
By dimensionless variables, $\bm{b}=P_{0}^{\infty}\bm{B}d_{r}/V_{s},\hat{\mathsf{G}}=\mathsf{G}/G,\hat{\dot{\bm{p}}}=\dot{\bm{p}}/G$, the equation for $P_0^\infty$ and $\bm{b}$ are written as
\begin{equation}
\nabla_{\bm{p}}\cdot(\Pe \dot{\bm{p}}P_{0}^{\infty}-\nabla_{\bm{p}}P_{0}^{\infty})=0~\textrm{with}~~
\int_{\mathbb{S}^2}P_{0}^{\infty}d^2\bm{p}=1
    \label{P0nondim},
\end{equation}
and
\begin{align}
    \nabla_{\bm{p}}\cdot[\text{Pe}\hat{\dot{\bm{p}}}\bm{b}-\nabla_{\bm{p}}\bm{b}]-\text{Pe}\bm{b}\cdot\hat{\mathsf{G}}=P_{0}^{\infty}(\bm{p}-\bar{\bm{p}})~\textrm{with}~~
    \int_{\mathbb{S}^2}\bm{b}d^2\bm{p}=0
    \label{nb},
\end{align}
respectively.
Similarly, the effective diffusion in \eqref{nDB} becomes
\begin{align}
    \bar{\mathsf{D}}=\frac{V_{s}^2}{d_{r}}\int_{\mathbb{S}^2}\left[\bm{b}\bm{p}+\text{Pe}\frac{\bm{b}\bm{b}}{P_{0}^{\infty}}\cdot\hat{\mathsf{G}}\right]^{\text{sym}}d^2\bm{p}
    \label{diffusion}.
\end{align}

To compute the partial differential equation on a unit sphere, Eqs.\eqref{P0nondim}--\eqref{nb}, we expand 
$P_{0}^{\infty}$ and $\bm{b}$ by spherical harmonics, with the polar and azimuthal angles $\theta$ and $\phi$ as in Fig. \ref{fig_micro}(a), via
\begin{align}
    P_{0}^{\infty}(\theta,\phi)&=\sum_{n=0}^{\infty}\sum_{m=0}^{n}(A_{n}^{m}\cos m\phi+B_{n}^{m}\sin m\phi)P_{n}^{m}(\cos\theta),
    \label{P_{0}}\\
    b_{j}(\theta,\phi)&=\sum_{n=0}^{\infty}\sum_{m=0}^{n}(\alpha_{nj}^{m}\cos m\phi+\beta_{nj}^{m}\sin m\phi)P_{n}^{m}(\cos\theta)\label{b},
\end{align}
where $P_{n}^{m}(x)$ is the associated Legendre polynomial. The coefficients $A_{n}^{m},B_{n}^{m},\alpha_{nj}^{m}$ and 
 $\beta_{nj}^{m}$ will be numerically computed. Hill \& Bees\cite{HB2002} computed the generalised Taylor dispersion for a spherical microswimmer with a truncation 
 $n_{\text{max}}=4$. For non-spherical swimmers and chiral swimmers, we need higher-order terms in the spherical harmonics to guarantee their numerical accuracy. In this study, we analysed by changing the series truncation from $n_{\text{max}}=10$ to  $n_{\text{max}}=30$, depending on the parameter sets.
Then the averaged velocity, $\bar{\bm{U}}$, are obtained as
\begin{equation}
\bar{\bm{U}}=\int_{\mathbb{S}^2}P_{0}^{\infty}(\bm{p})\bm{U}_s(\bm{p})d^2\bm{p} 
=\frac{4\pi}{3}V_s\
\left( A_{1}^{1}, B_{1}^{1}, A_{1}^{0}\right)^{\textrm{T}}
    ,\label{Us}
\end{equation}
%\begin{align}
%    \bar{\bm{U}}&=\int_{\mathbb{S}^2}P_{0}^{\infty}(\bm{p})\bm{U}_s(\bm{p})d^2\bm{p} \nonumber \\ 
%&=
%    \frac{4\pi}{3}V_s\begin{pmatrix}     A_{1}^{1} \\ B_{1}^{1} \\ A_{1}^{0}     \end{pmatrix} 
%+\frac{2\pi}{5}\beta \begin{pmatrix}     2B_2^{2} \\ A_{0}^{2}-12A_{2}^{2} \\ -B_{2}^{1}    \end{pmatrix}
%    +\frac{2\pi}{3}\gamma   \begin{pmatrix}    A_1^{0} \\ 0 \\ A_{1}^{1}    \end{pmatrix}
%    +\frac{4\pi}{105}(\delta-\gamma)    \begin{pmatrix}    -3A_3^{0}+7A_1^{0}+30A_{3}^{2} \\ 30B_{3}^{2} \\ 12A_{3}^{1}+7A_{1}^{1}    \end{pmatrix}    ,\label{Us}
%\end{align}
and 
the effective diffusion tensor in the absence of the background flow is computed as 
\begin{equation}
    \bar{\mathsf{D}}=\frac{2\pi}{3}\frac{V_{s}^2}{d_{r}}\int_{\mathbb{S}^2}[\bm{b}\bm{p}]^{\text{sym}}d^2\bm{p}=\frac{2\pi}{3}\frac{V_{s}^2}{d_{r}}\begin{pmatrix}
    2\alpha_{11}^{1} & \alpha_{12}^{1}+\beta_{11}^{1} & \alpha_{11}^{0}+\alpha_{13}^{1}\\
    \alpha_{12}^{1}+\beta_{11}^{1} & 2\beta_{12}^{1} & \alpha_{12}^{0}+\beta_{13}^{1}\\
    \alpha_{11}^{0}+\alpha_{13}^{1} & \alpha_{12}^{0}+\beta_{13}^{1} & 2\alpha_{13}^{0}
    \end{pmatrix}
\label{Dbar}.
\end{equation}
The correction term in the diffusion tensor due to the background shear, the second term of \eqref{diffusion}, will be computed by a direct numerical integration over a unit sphere.

To obtain the values of the coefficients, we  proceed to set up the equations to solve Eqs. \eqref{P0nondim}--\eqref{nb}. Substituting the expression for the background flow field $\bm{V}=(Gz,0,0)^{\text{T}}$ leads to the angular dynamics of the chiral particle, with a non-dimensionalised gyrotactic timescale $g=1/BG$, given by
\begin{align}
    \dot{\theta}&=-\frac{g}{2}\sin\theta+\frac{1}{2}(1+b\cos 2\theta)\cos\phi-\frac{c}{2}\cos\theta\sin\phi,
    \label{theta}\\
    \dot{\phi}\sin\theta&=-\frac{1}{2}(1+b)\cos\theta\sin\phi-\frac{c}{2}\cos 2\theta\cos\phi \label{phi}.
\end{align}
The equations for $P_{0}^{\infty}$ and $\bm{b}$, Eqs. \eqref{P0nondim}--\eqref{nb}, are then summarised as
\begin{equation}
    \mathcal{L} P_{0}^{\infty}=0,
    ~\textrm{and}~~
    \mathcal{L} b_{j}-\text{Pe}b_{3}\delta_{1j}=P_{0}^{\infty}(p_{j}-\bar{p}_{j})\quad (j=1,2,3)
    \label{beq},
\end{equation}
where the linear operator $\mathcal{L}$ is defined as
\begin{align}
\begin{split}
    \mathcal{L}(\cdot)=\frac{\text{Pe}}{2}&\left[\{-g\sin\theta+(1+b\cos 2\theta)\cos\phi-c\cos\theta\sin\phi\}\pdv{\theta}(\cdot)-\{(1+b)\cos\theta\sin\phi+c\cos 2\theta\cos\phi\}\right.\\
    &\frac{1}{\sin\theta}\pdv{\phi}(\cdot)-(2g\cos\theta+6b\sin\theta\cos\theta\cos\phi)(\cdot)\biggl]
    -\frac{1}{\sin\theta}\pdv{\theta}\left(\sin\theta\pdv{\theta}(\cdot)\right)-\frac{1}{\sin^2\theta}\pdv[2]{\phi}(\cdot)
\end{split}
\label{L}.
\end{align}

We further expand this set of equations with the spherical harmonics and rewrite the system into the set of linear relations between the coefficients of the spherical harmonics. Some details are shown in the Appendix \ref{app}.

\section{Orientational dynamics and its distribution}
\label{sec:ori}

\subsection{Individual deterministic dynamics}

Before proceeding to the numerical results of population-level distribution, it is beneficial to examine the deterministic angular dynamics described by Eqs, \eqref{theta}-\eqref{phi}.
The fixed points of the dynamics, where $\dot{\theta}=\dot{\phi}=0$, are readily obtained as
\begin{equation}
    \tan\phi=-\frac{c\cos 2\theta}{(1+b)\cos\theta} ~\textrm{and}~~
     f_{\pm}(\theta)=0,
\end{equation}
where $f_{\pm}(\theta)$ are given by
\begin{align}
    f_{\pm}(\theta)=-g\sin\theta\left[(1+b)^2\cos^2\theta+c^2\cos^2 2\theta\right]^{1/2}\pm\cos\theta\left[c^2\cos 2\theta+(1+b)(1+b\cos 2\theta)\right]
    \label{fpm}.
\end{align}
If the gyrotactic effects are negligible ($g=0$), these fixed points are simply represented. When $b^2+c^2\leq 1$, there are only two fixed points, given by $(\phi,\theta)=\left(\pm \pi/2,\pi/2\right)$. When $b^2+c^2>1$, however, there are 6 fixed points, 
\begin{align}
    (\phi,\theta)=\left(\pm\frac{\pi}{2},\frac{\pi}{2}\right),\quad(\phi_{1},\theta_{1}),\quad(\phi_{1}-\pi,\theta_{1}),\quad(-\phi_{1},\pi-\theta_{1}),\quad(\pi-\phi_{1},\pi-\theta_{1}),
\end{align}
where $\phi_{1}$ and $\theta_{1}$ are given by
\begin{equation}
    \phi_{1}=\arctan\sqrt{\frac{2c^2}{(b^2+b+c^2)(b^2+c^2-1)}} ~\textrm{and}~~
    \theta_{1}=\arccos\sqrt{\frac{b^2+c^2-1}{2(b^2+b+c^2)}},
\end{equation}
respectively.

\begin{figure}[!t]
\centering
%\begin{overpic}[width=6cm]{flowline/flowline_g=0_b=0.5.png}
%\put(10,66){(a) $g=0,b=0.5,c=0$}
%\end{overpic}
%\begin{overpic}[width=6cm]{flowline/flowline_g=0_b=0.5_c=0.3.png}
%\put(-5,60){(b) $g=0,b=0.5,c=0.3$}
%\end{overpic}
\begin{overpic}[width=6cm]{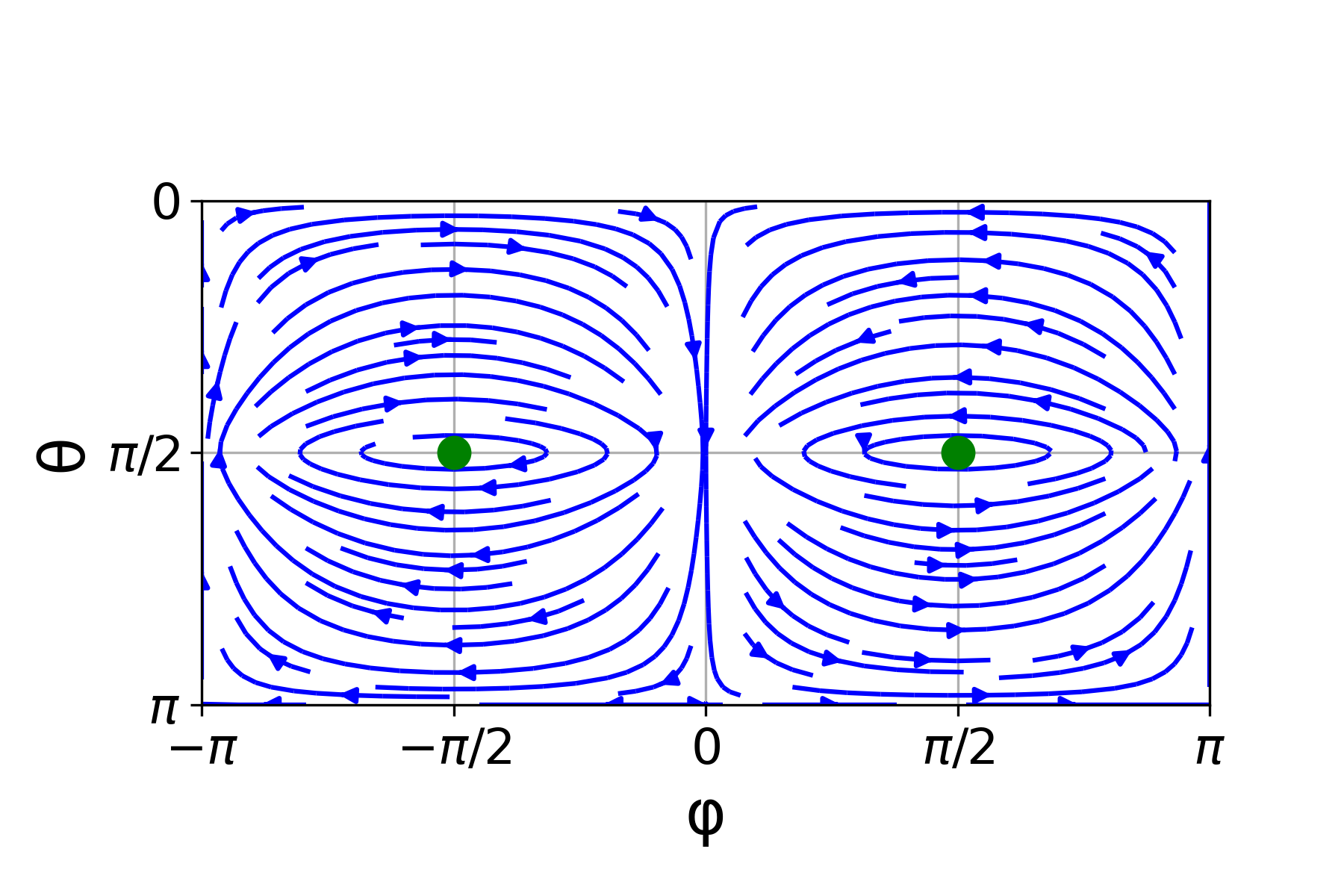}
\put(20,60){(a) ~$g=0,b=0.95,c=0$}
\end{overpic}
\begin{overpic}[width=6cm]{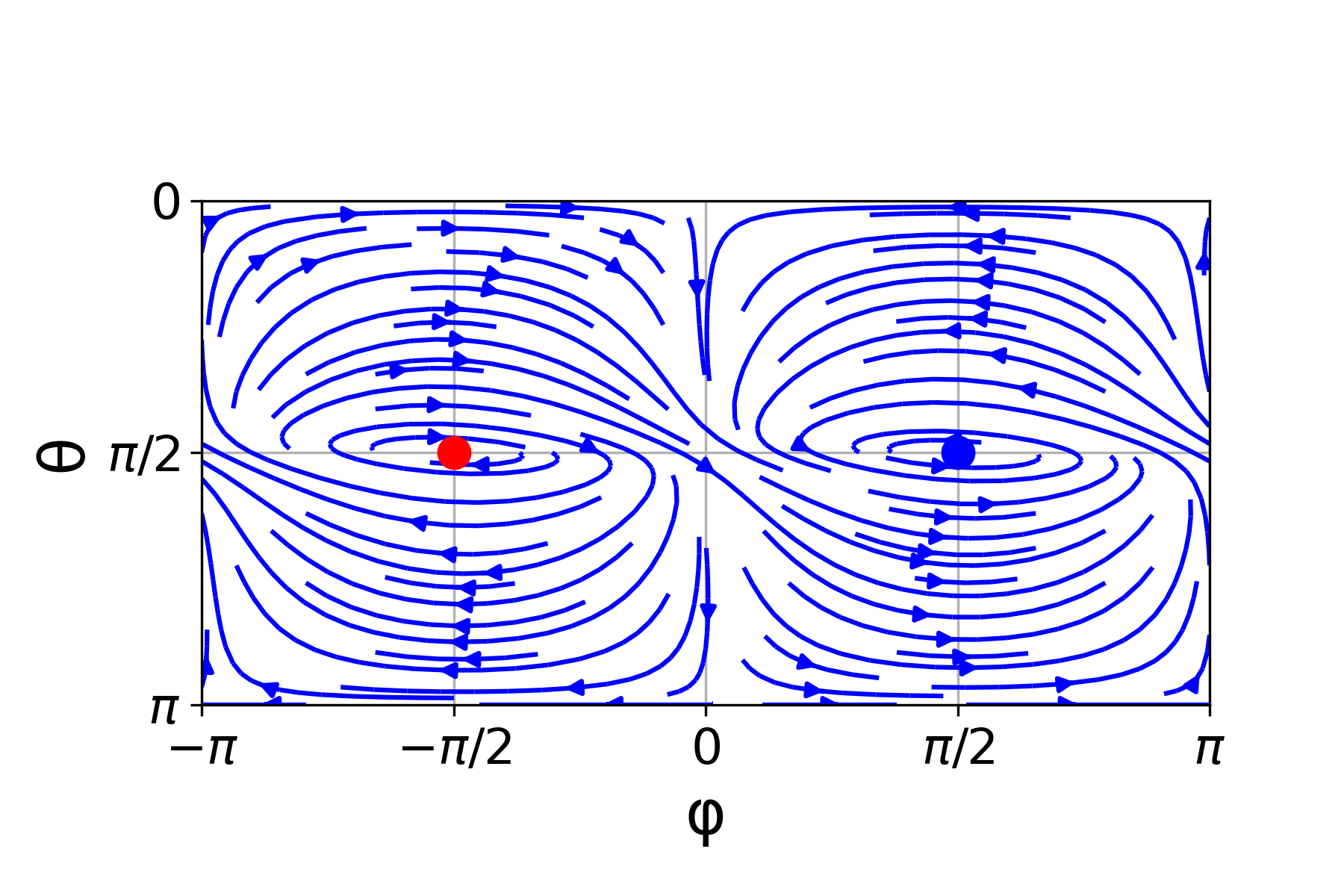}
\put(20,60){(b) ~$g=0,b=0.95, c=0.1$}
\end{overpic}
%\begin{overpic}[width=6cm]{flowline/flowline_b=0.95_c=1.0.png}
%\put(20,60){(c) ~$g=0,b=0.95,c=1.0$}
%\end{overpic}
\begin{overpic}[width=6cm]{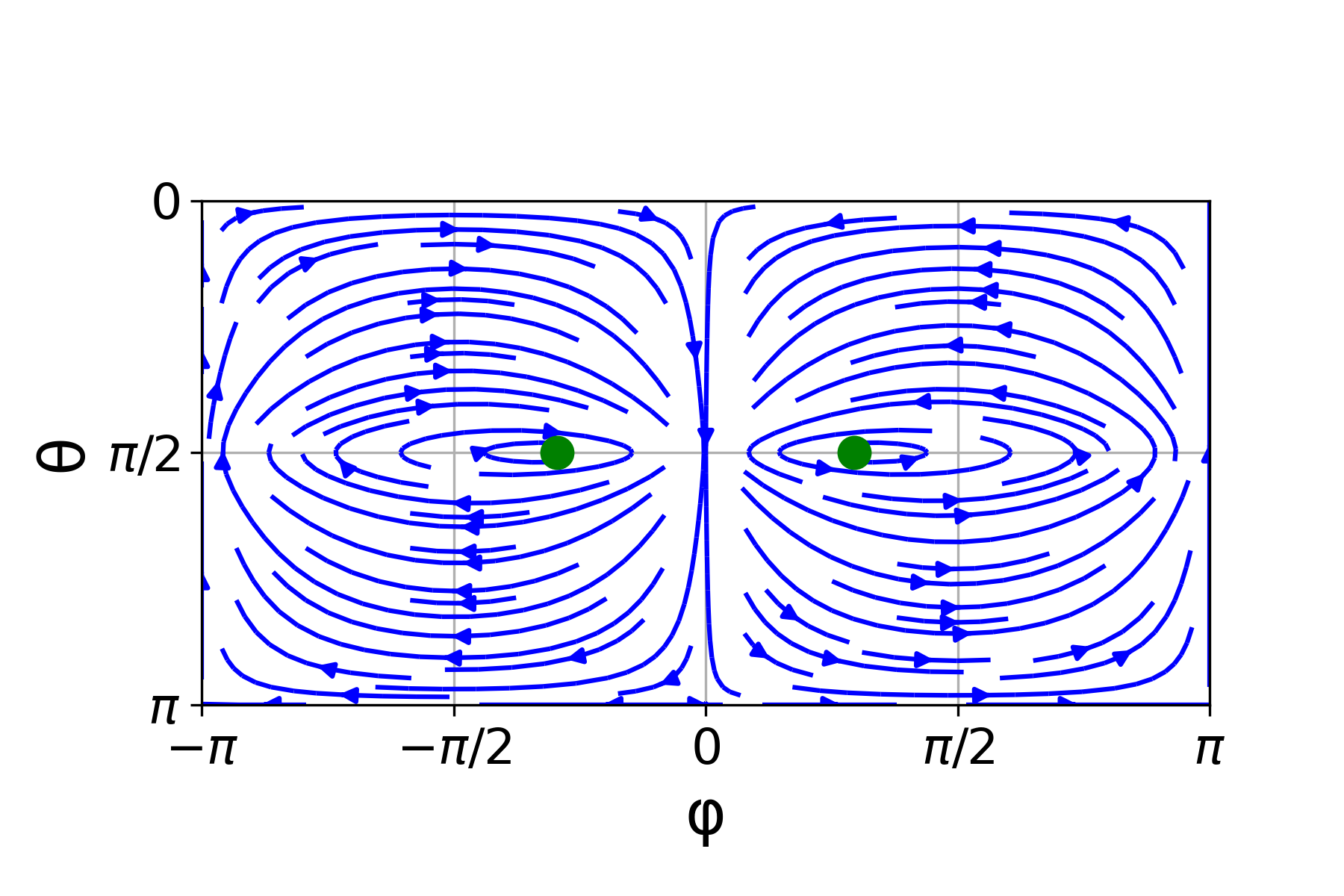}
\put(20,60){(c) ~$g=0.03,b=0.95,c=0$}
\end{overpic}
%\begin{overpic}[width=6cm]{flowline/flowline_g=0.2_b=0.95.png}
%\put(20,60){(e) ~$g=0.2,b=0.95,c=0$}
%\end{overpic}
\begin{overpic}[width=6cm]{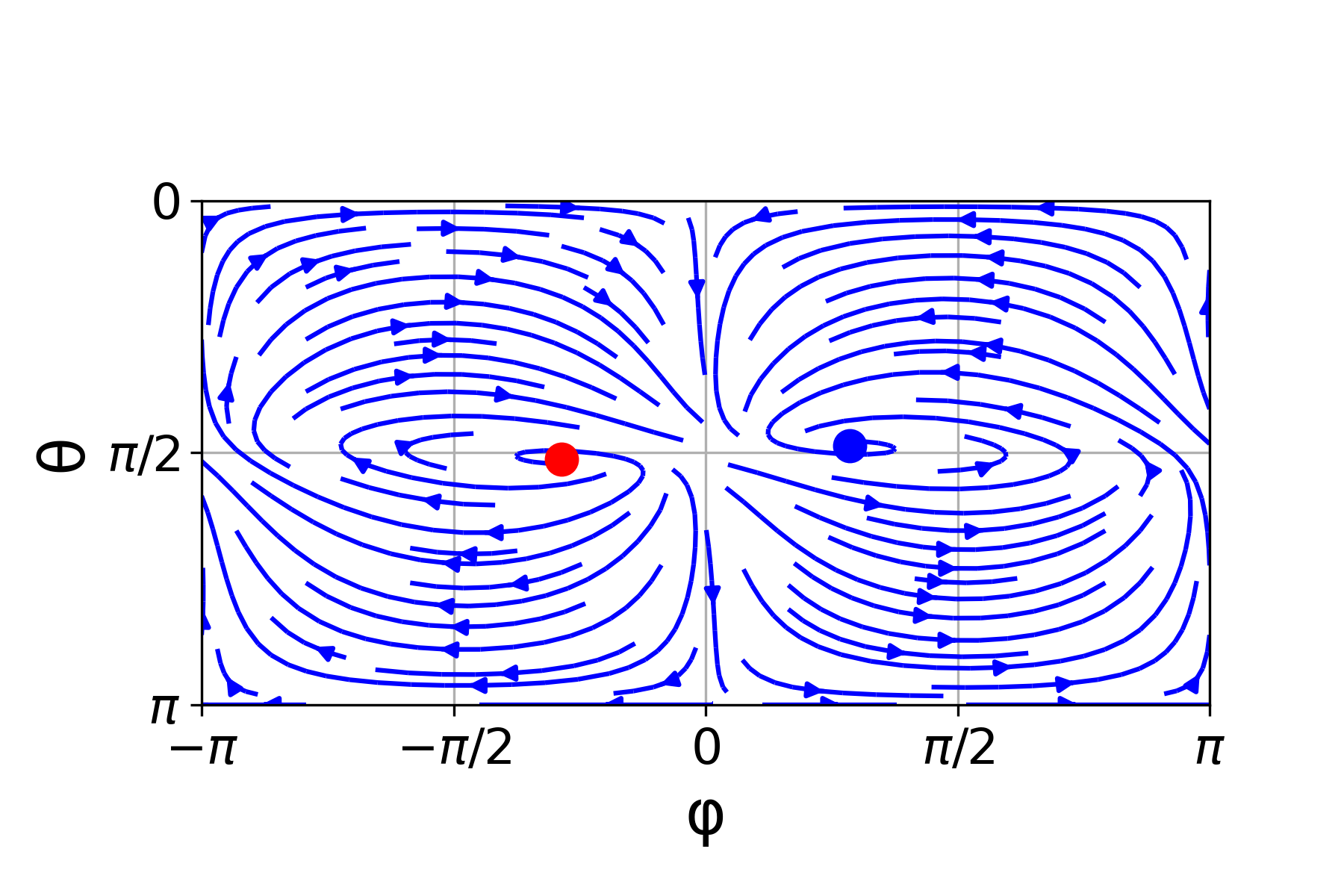}
\put(20,60){(d) ~$g=0.03,b=0.95, c=0.1$}
\end{overpic}
    \caption{Phase portrait of the generalised Jeffery equation for a chiral particle under a gyrotactic torque ($g$) with different shape parameters, $b$ and $c$. The streamlines in the $\phi-\theta$ plane are shown and the fixed points are marked with different colours and symbols (blue circle: attractive fixed point, red circle: repelling fixed point, green circle: neutrally stable, and green diamond: saddle fixed point). (a) A periodic motion known as the Jeffery orbit. (b) attractive and repelling fixed points emerge due to the chirality ($b^2+c^2<1)$. %(c) New fixed points bifurcates when $b^2+c^2$ exceeds one. 
    (c) The gyrotactic torque only modulates the periodic orbits at a small $g$ with $g<1-b$. %(e) A stable and unstable pair emerges at a large $g$ with $g>1-b$. 
    (d) The gyrotactic torque also modulates the position of attracting and repelling fixed points for a chiral particle.}
    \label{fl}
  \end{figure}

In Fig.\ref{fl}, we present streamlines and the stability of these fixed points with different gyrotactic and shape parameters. Throughout the paper, to focus on biologically relevant situations, we will consider $g=0.03$ for typical gyrotactic time scale as seen in Chlamydomonas \cite{JC2020,Bees2020}, and for shape parameters we will consider $b=0.95$ and $c=0.1$ for typical chiral microswimmers, following recent experiments of E. coli bacteria by Jing et al. \cite{Jing2020}. %, where their corresponding parameters are $g = 0,b = 0.923,c = 0.06$. 

For a non-chiral particle ($c=0$), there are two fixed points in the angular dynamics. By analysing the zeros of $f_{\pm}$ in Eq. \eqref{fpm}, we found when $g>1-b$ that one of the fixed points located in $0<\theta<\pi/2$ is linearly stable and the other fixed point is linearly unstable. When $g\leq 1-b$, these two fixed points become neutrally stable.
As shown in Fig. \ref{fl}(a,b), a chiral particle with $c>0$ and $g=0$ possesses one attractive and one repelling fixed point in the phase portrait, while the non-chiral particle ($g=c=0$) follows the periodic motions of the Jeffery orbit. The linear stability analysis of these fixed points leads to zero real parts of the eigenvalues and attraction and repelling dynamics are known to be slower than exponential in time and scales as $t^{-1/2}$ \cite{Ishimoto2020}. 
In many biological swimmers, the gyrotactic torque is sufficiently weak to satisfy $g<1-b$ and only found to modulate the angular dynamics to attract towards $\phi=0$, which corresponds to the flow direction [Fig.\ref{fl}(c,d)].

\subsection{Orientational distribution and averaged swimming velocity}
%\begin{itemize}
%    \item Figure 5.1. $P_0^\infty$
%    \item At large $\Pe$, the distribution results are interpreted from the dynamical systems of the (chiral) Jeffery equation
%    \item At small $\Pe$, the distribution follows the $\nabla_{\bm{p}}\cdot \dot{\bm{p}}$ determines the orientational distribution.
%    \item Our simulation results are in reasonable agreement with experiments and stochastic simulation presented in Jing et al. \cite{Jing2020}, where the rotational diffusion was set as $d_r = 0.057$sec$^{-1}$ for E.coli bacteria and hence $\Pe = G/d_r \approx 17.54 \times  G$. Jing et al. \cite{Jing2020} present the simulation results with $G=1, 10, 100, 1000$sec$^{-1}$.
%\end{itemize}

We now proceed to present our numerical results on the probability distribution of the particle orientation $P_{0}^{\infty}(\theta,\phi)$.

\begin{figure}[t!]
\centering
\begin{overpic}[width=6.5cm]{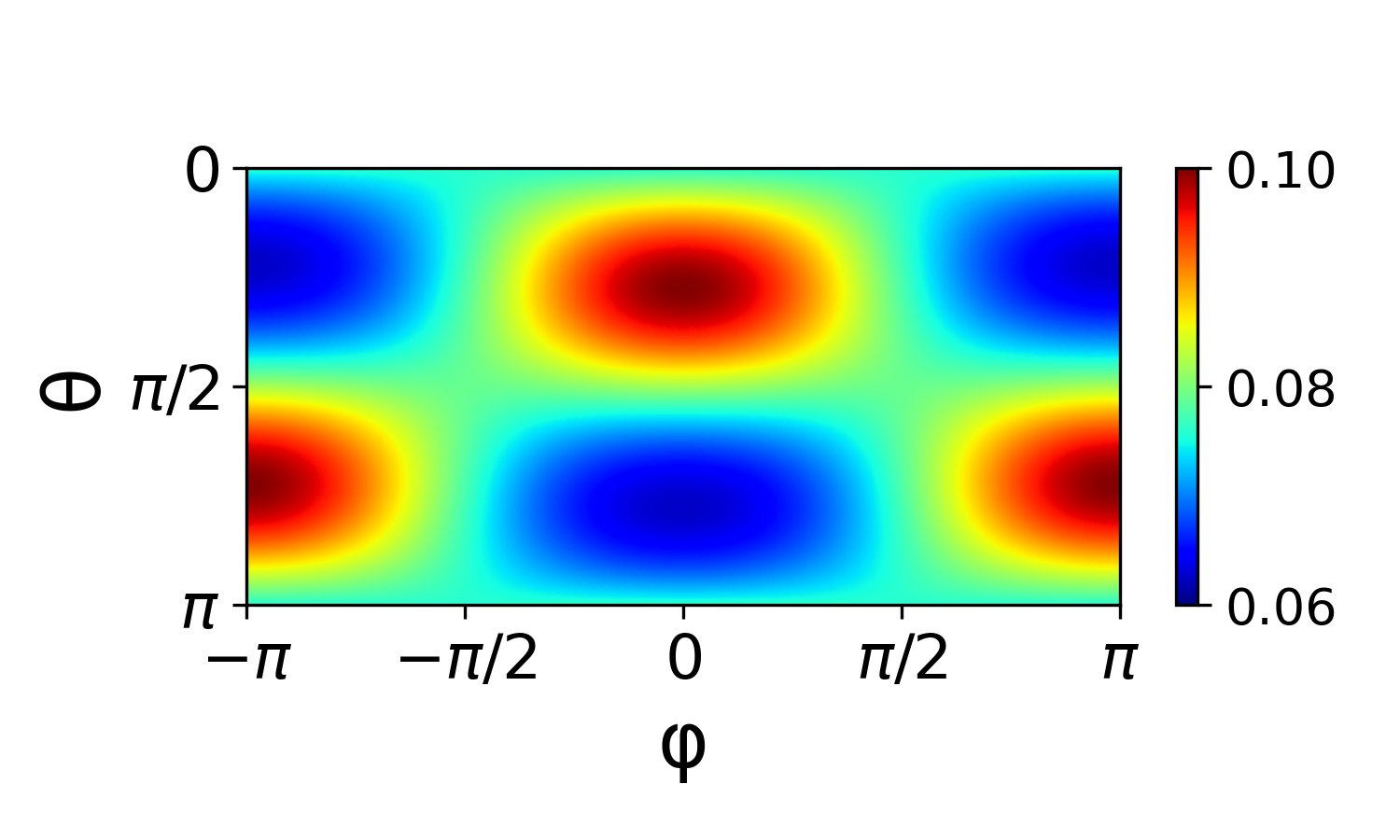}
\put(13,52){(a) ~$g=0,b=0.95,c=0.1, \Pe=1$}
\end{overpic}
\begin{overpic}[width=6.5cm]{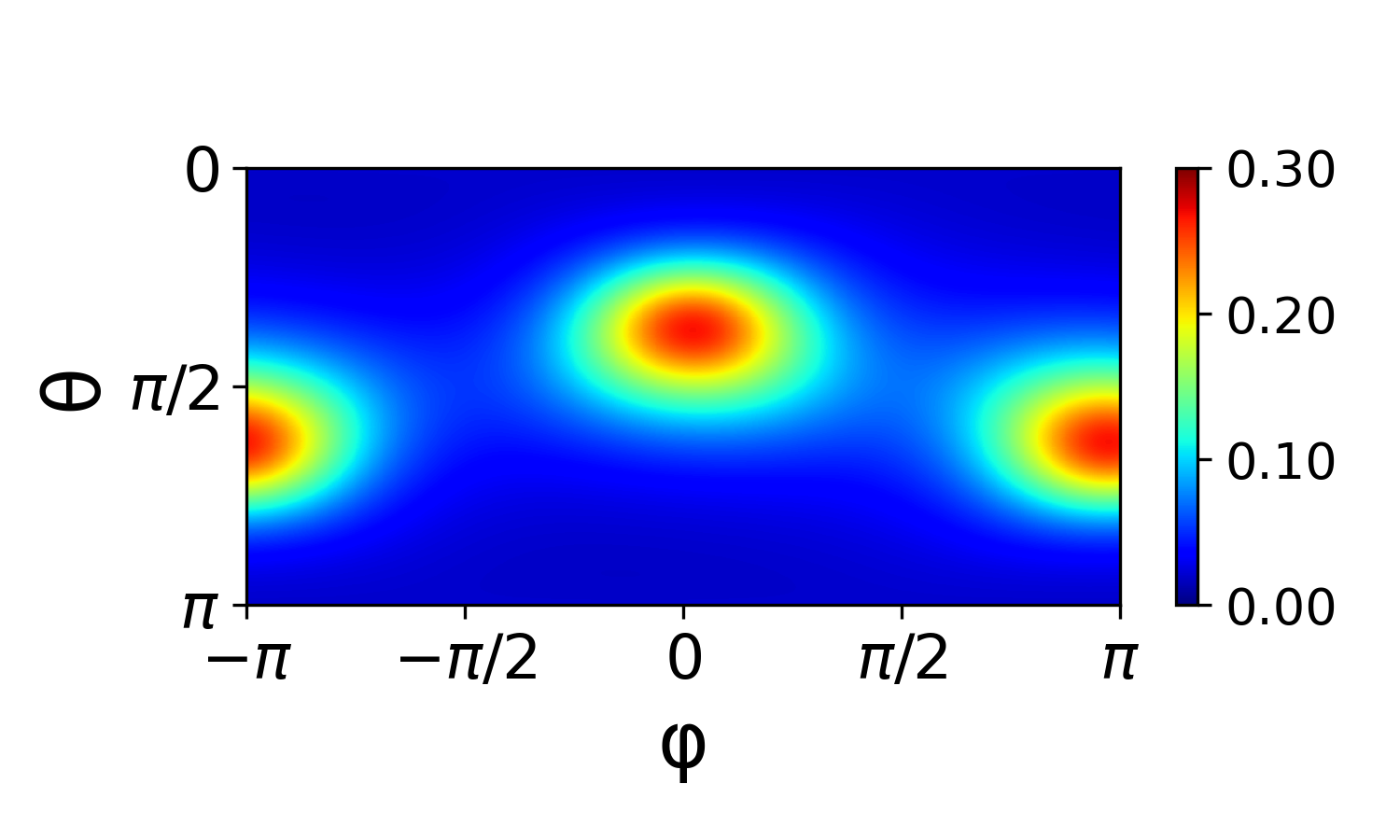}
\put(13,52){(b) ~$g=0,b=0.95,c=0.1, \Pe=10$}
\end{overpic}
%\begin{overpic}[width=6.5cm]{P0/P0_g=0_b=0.95_c=0.1_n=28_Pe=200.png}
%\put(15,52){(c) ~$g=0,b=0.95,c=0.1$}
%\end{overpic}
    \caption{Probability distribution of the orientation $P_{0}^{\infty}(\theta,\phi)$ at (a) $\Pe=1$ and (b) $\Pe=10$ for chiral particles with $g=0, b=0.95, c=0.1$. At a small $\Pe$, the distribution is almost symmetric to $\phi=0$ and the chiral effects are irrelevant}
    \label{P0}
  \end{figure}
  
When $\Pe$ is small, the distribution is almost uniform in the $\theta-\phi$ space. By expanding the series
\begin{equation}
    P_0^\infty=\frac{1}{4\pi}+\Pe P_1^\infty + \Pe^2 P_2^\infty +\cdots
\end{equation}
and plugging into Eq.\eqref{P0nondim}, the $O(\Pe)$ term reads the Poisson equation on an unit sphere as
\begin{equation}
    \nabla_{\bm{p}}^2 P_1^\infty=\frac{1}{4\pi}\nabla_{\bm{p}}\cdot\dot{\bm{p}}=\frac{1}{4\pi}\left(-g\cos\theta-3b\sin\theta\cos\theta\cos\phi\right),
\end{equation}
where the source term on the right-hand side is independent of the chirality effects. This chirality independence is also seen in Fig. \ref{P0}, where we show the probability distribution $P_0^\infty$ at small $\Pe$ ($\Pe=1$ and $\Pe=10$). As in the figure, the probability distribution is almost symmetric to $\phi=0$ even for chiral particles.

\begin{figure}[t!]
\centering
\begin{overpic}[width=6.5cm]{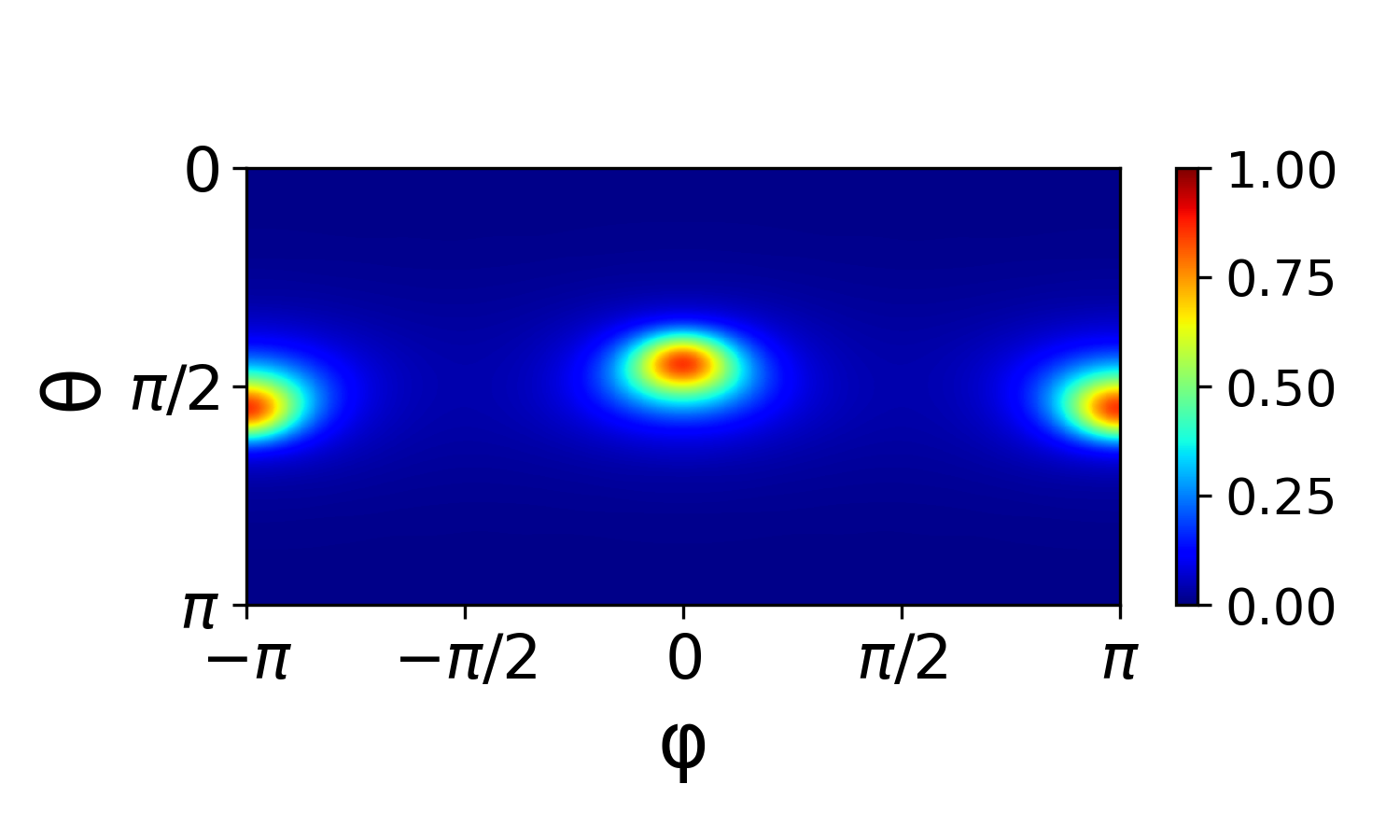}
\put(19,52){(a) ~$g=0,b=0.95,c=0$}
\end{overpic}
\begin{overpic}[width=6.5cm]{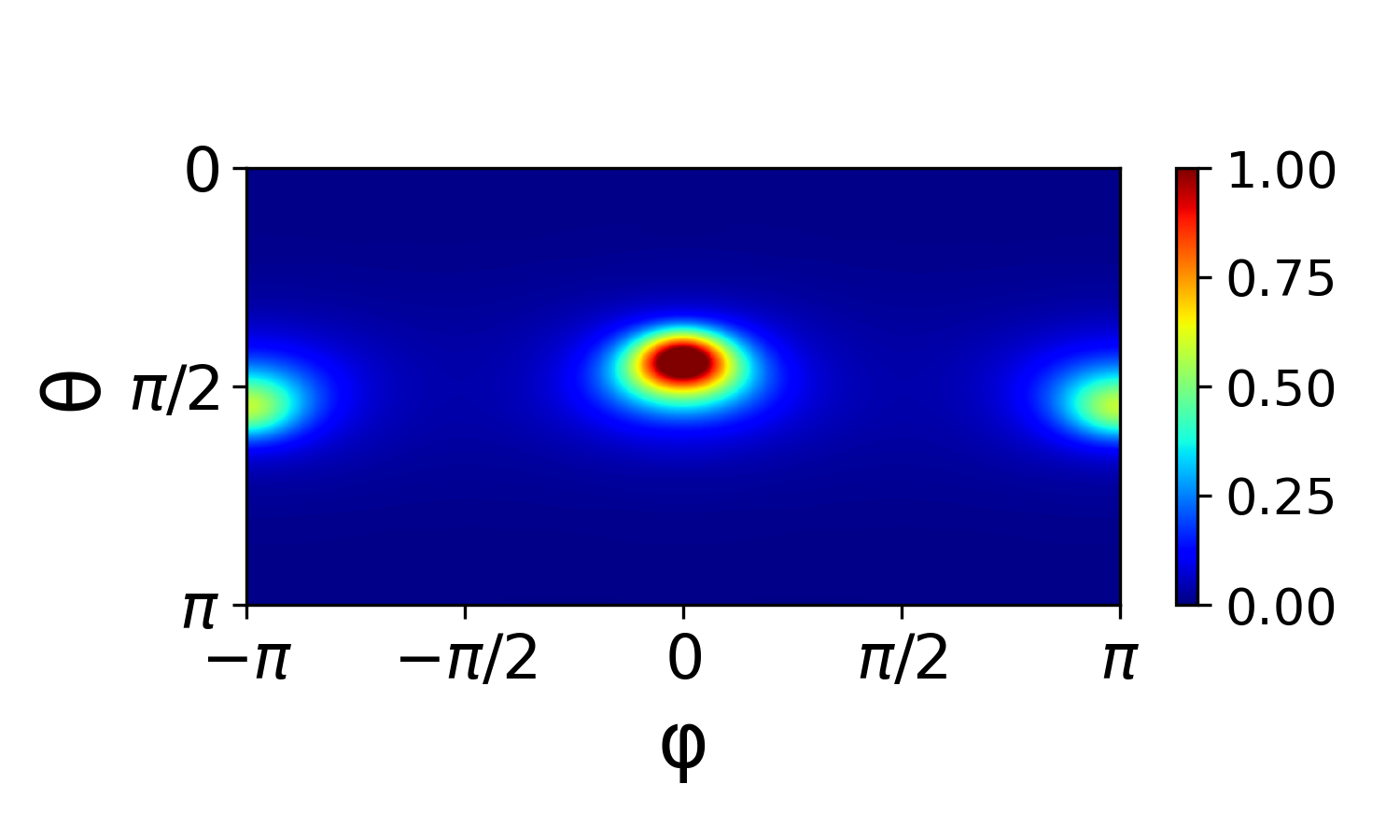}
\put(19,52){(b) ~$g=0.03,b=0.95,c=0$}
\end{overpic}
\begin{overpic}[width=6.5cm]{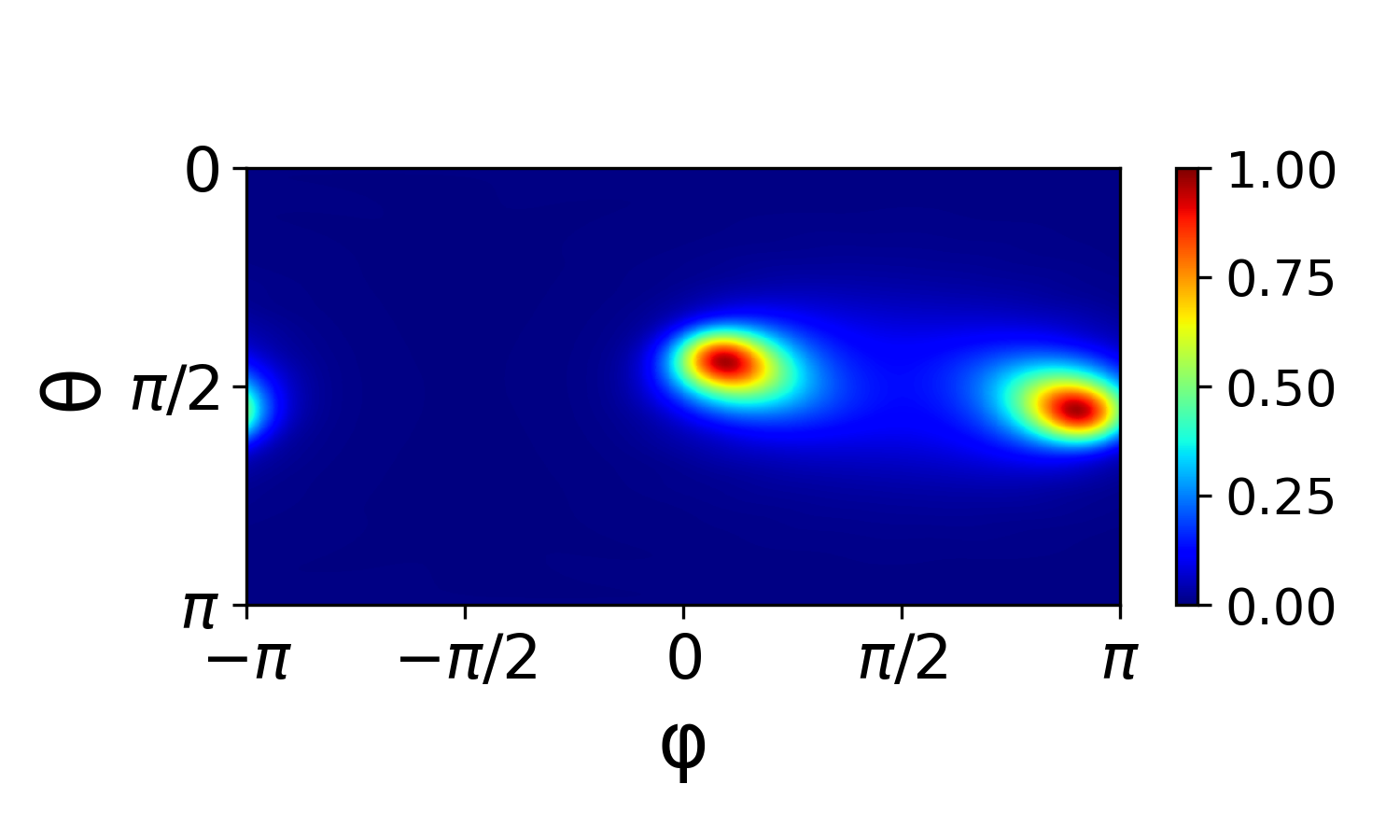}
\put(19,52){(c) ~$g=0,b=0.95,c=0.1$}
\end{overpic}
\begin{overpic}[width=6.5cm]{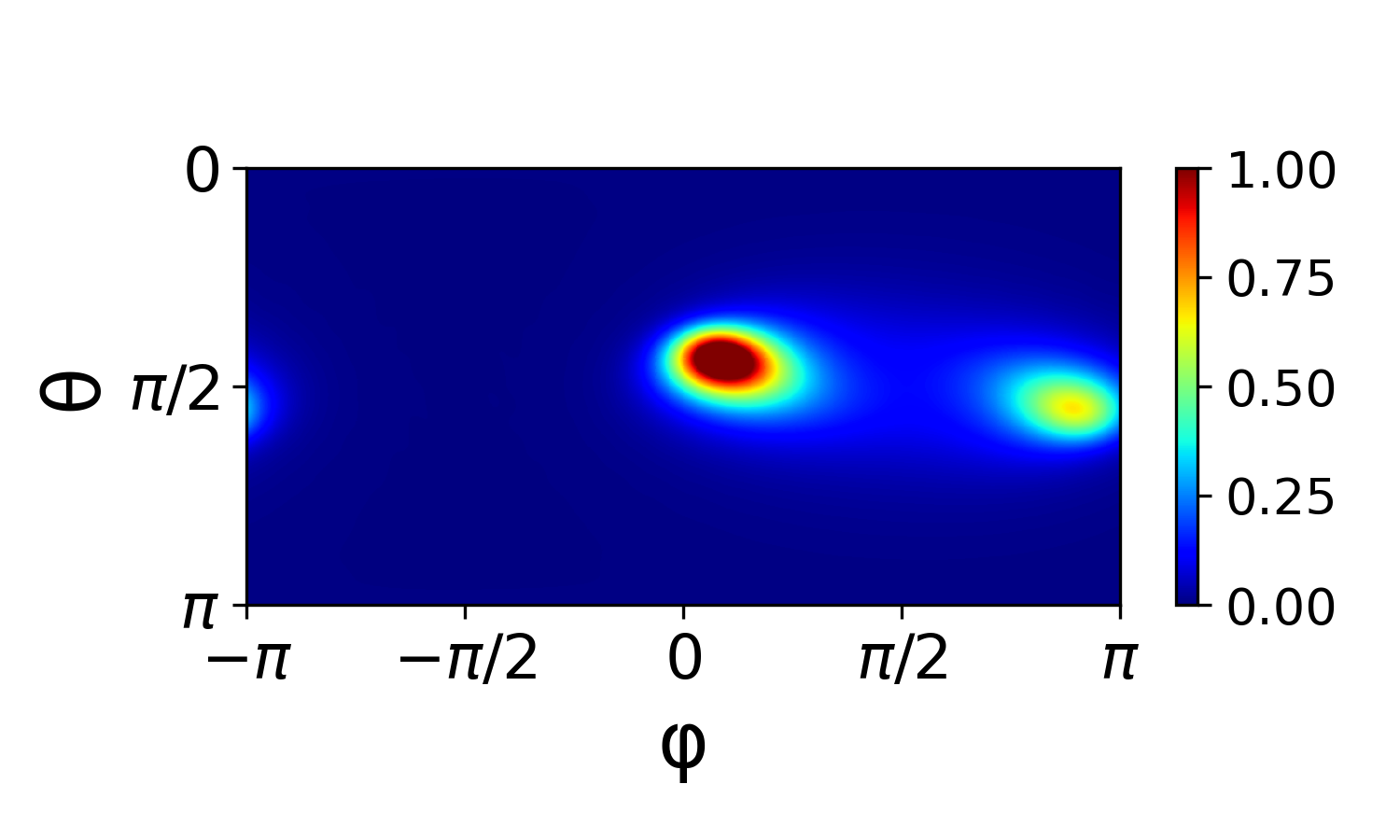}
\put(19,52){(d) ~$g=0.03,b=0.95,c=0.1$}
\end{overpic}
    \caption{Probability distribution of the particle orientation $P_{0}^{\infty}(\theta,\phi)$ at $\Pe=100$ with different gyrotactic and shape parameters.  (a) The particle orientation is symmetrically localised at $\phi=0,\pi$ in the absence of gyrotactic and chirality effects. (b) If only the gyrotactic torque is applied $g>0$, the orientation becomes biased at $\phi=0$. (c) If only the chirality effect is considered, the particle orientation is bimodally distributed around $\phi=\pi/2$. (d) When both the gyrotactic and chirality effects are taken into consideration, the distribution becomes localised around one of the peaks near $\phi=0$.}
    \label{P0c}
  \end{figure}

In Fig. \ref{P0c}, we show a colour map of $P_{0}^{\infty}(\theta,\phi)$ in colour with different gyrotactic and shape parameters at $\Pe= 100$. When the particles are non-chiral, the orientation distribution is still symmetric to $\phi=0$ at the higher value of $\Pe$ as in the deterministic dynamics in Fig.\ref{fl}.
When the particle is chiral, however, there is an attractive fixed point in the $\theta-\phi$ phase space, which breaks the symmetry to $\phi=0$. Hence, as the $\Pe$ increases, the distribution of the orientation of a chiral particle accumulates at the attracting fixed point in the $\theta-\phi$ space [Fig.\ref{fl}]. Nonetheless, since the attraction is weaker than an exponential function in time, even with a large $\Pe$, the orientation probability is distributed around the attracting points as seen in Fig. \ref{P0c}(c, d). 

In particular, when $g=0$ a chiral particle exhibits a bimodally-distributed orientational probability around $\phi=\pi/2$ [Fig. \ref{P0c}(c)].
This result is in agreement with experiments and stochastic simulation by Jing et al. \cite{Jing2020}, where the rotational diffusion was set as $d_r = 0.057$sec$^{-1}$ for E. coli bacteria and hence $\Pe = G/d_r \approx 17.54 \times  G$. Jing et al. presents the simulation results with $G=1, 10, 100, 1000$sec$^{-1}$ and a bimodal distribution at an intermediate shear strength with $G=10$sec$^{-1}$.

The gyrotactic torque ($g>0$) attracts the orientation around $\phi=0$ as predicted from the analysis of the deterministic dynamical systems. Fig.\ref{P0c}(d) shows that the distribution becomes localised at the left peak near $\phi=0$ from the bimodal distribution of Fig.\ref{P0c}(c).

%\subsection{Averaged swimming velocity}
%\begin{itemize}
%    \item Figure 5.2. $p_x$, $p_y$, $p_z$
%    \item {\rd [We will present the orientation vector]}
%\end{itemize}

\begin{figure}[t!]
\centering
%\begin{overpic}[width=4cm]{pxpypz/chiral_pxpz_g=0.03_b=0.5_c=0_n=10_integral.png}
%\put(15,68){(a) ~$g=0.03,b=0.5,c=0$}
%\end{overpic}
\begin{overpic}[width=4.4cm]{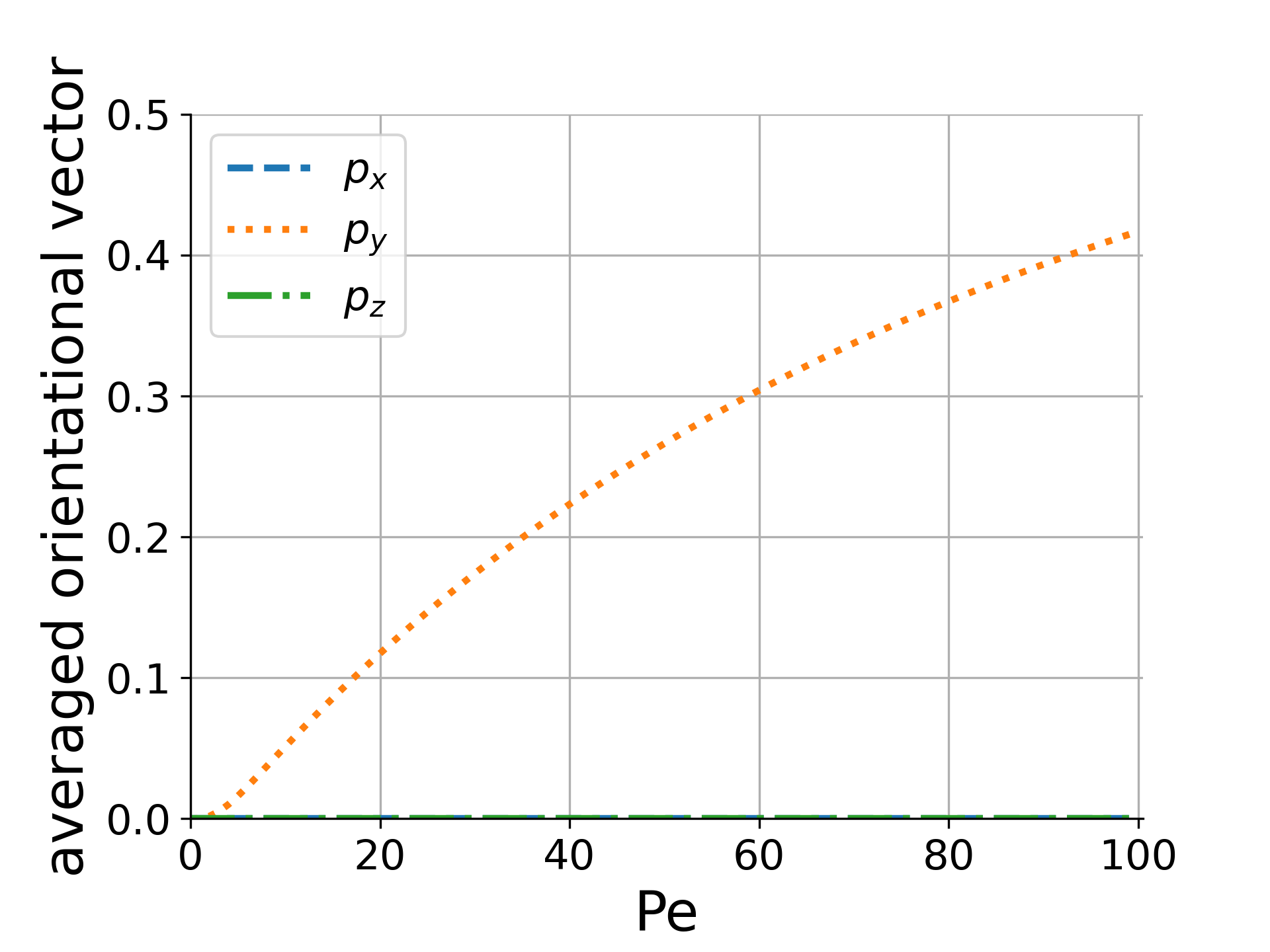}
\put(7,71){(a) ~$g=0,b=0.95, c=0.1$}
\end{overpic}
\begin{overpic}[width=4.4cm]{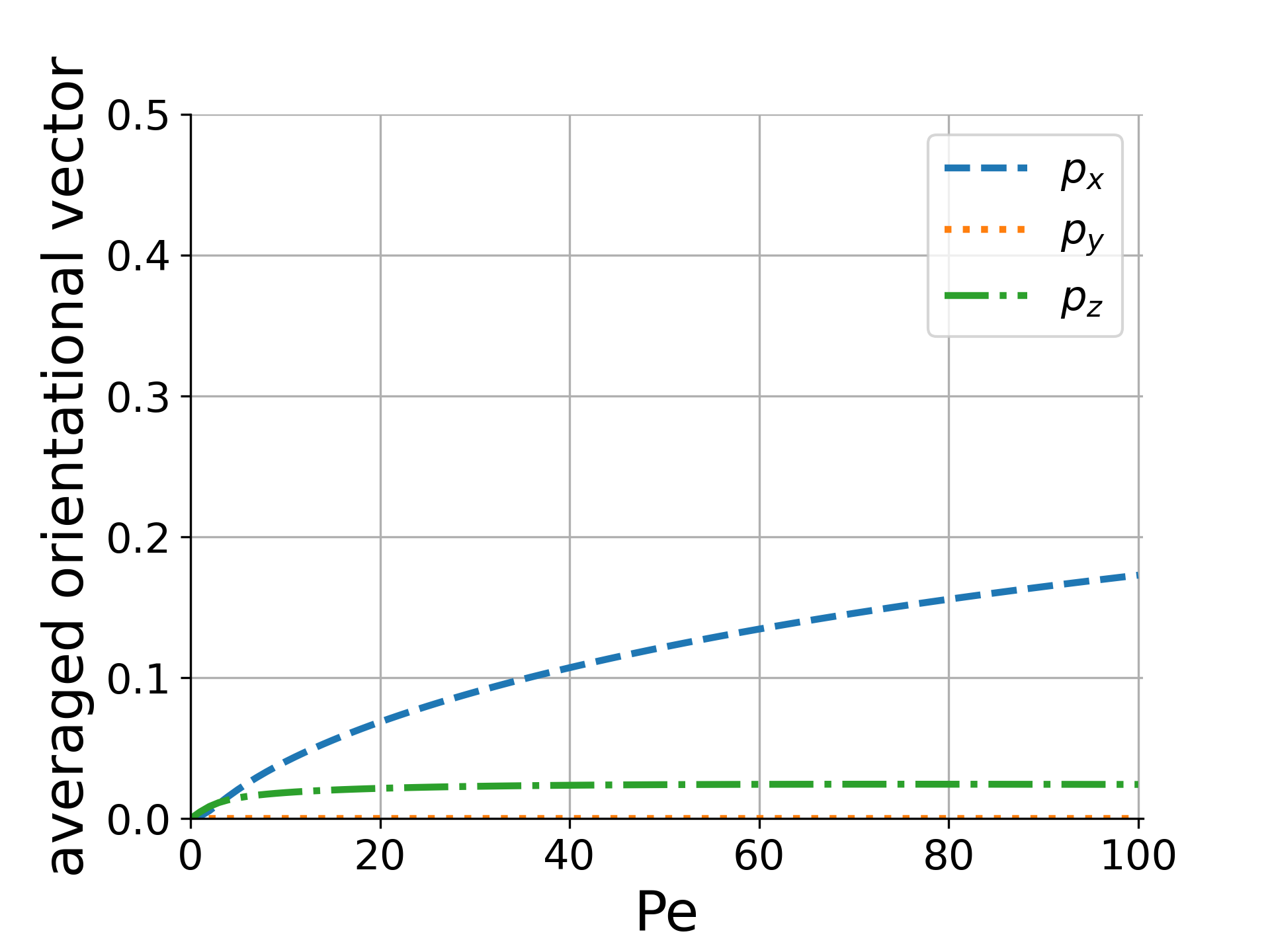}
\put(7,71){(b) ~$g=0.03,b=0.95,c=0$}
\end{overpic}
\begin{overpic}[width=4.4cm]{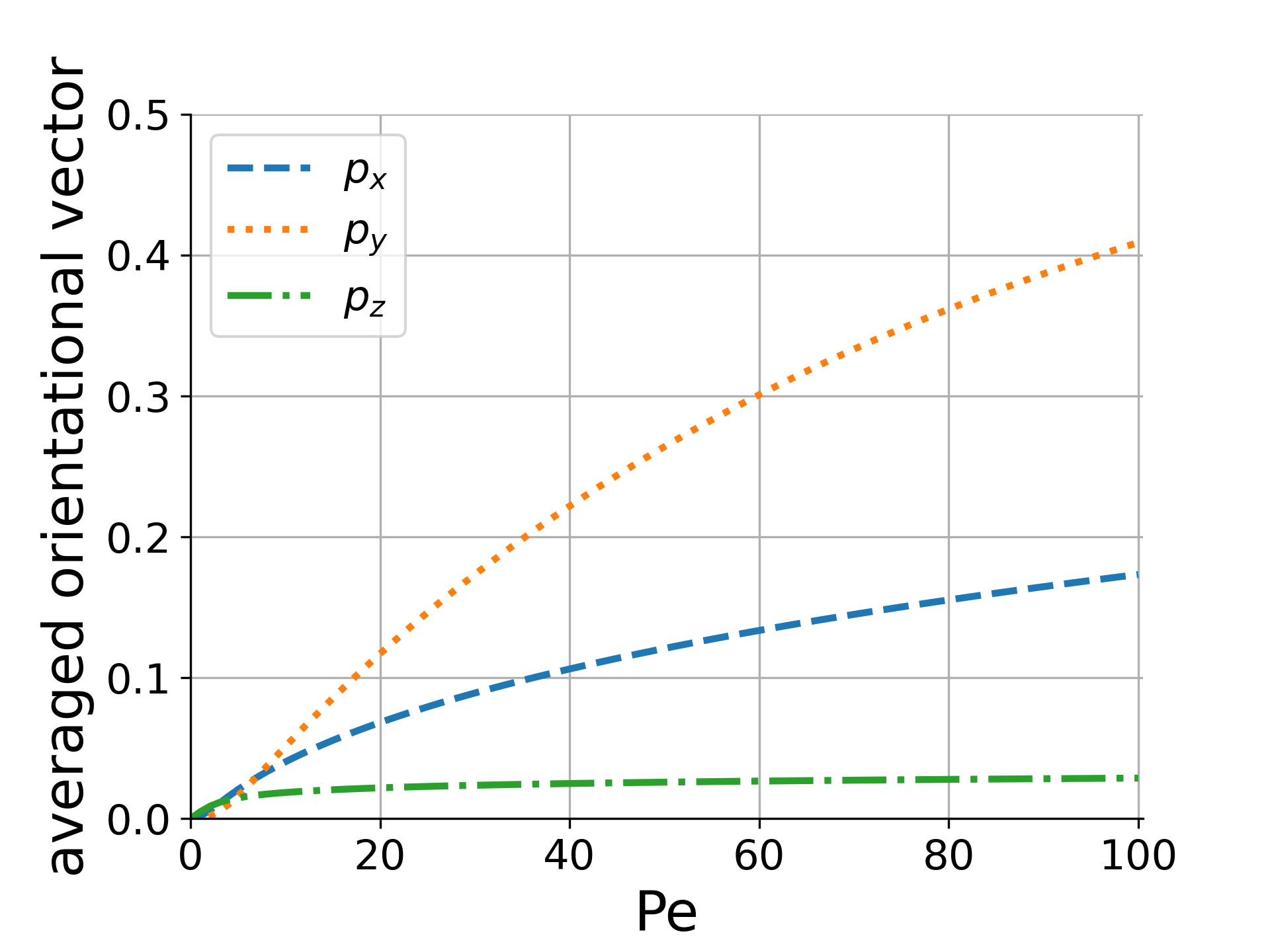}
\put(7,71){(c) ~$g=0.03,b=0.95, c=0.1$}
\end{overpic}
    \caption{Components of averaged orientation $p_{x},p_{y},p_{z}$ with different $\Pe$ with $0\leq\text{Pe}\leq 100$. (a) A chiral particle with $c>0$ directs towards the background vorticity vector ($p_y>0$), while the symmetry in $x$ and $z$ axis results in zero-mean components, irrespective of the value of $\Pe$.
    (b) The gyrotactic torque breaks the symmetry in $x$ and $z$ axes but holds the symmetry in $y$ axis, yielding the non-zero mean values only for $p_x$ and $p_z$. (c) The effects of gyrotactic torque and chiral torque exhibit the symmetry breaking to generate non-zero mean values in each component almost independently.}
    \label{pxpypz}
  \end{figure} 
  
We then compute the averaged orientation $\bar{\bm{p}}$ with different $\Pe$ and show the plots in Fig.\ref{pxpypz}. From Eq. \eqref{Us}, this readily provides the averaged swimming velocity by $\bar{\bm{U}}=(4\pi/3)V_s\bar{\bm{p}}$.%, if $\beta=\gamma=\delta=0$.These shape parameters indeed vanish for a symmetrical particle called the heterochiral objects \cite{Ishimoto2020}. Further, these drift velocities due to the shape parameters are negligibly small compared with the swimming velocity for typical bacterial cells. Hence, in this section, we focus on the case with $\beta=\gamma=\delta=0$, which allows us to interpret the averaged orientation vector as the averaged swimming velocity.

The plots for a chiral microswimmer $(c>0)$ without gyrotactic torque show a non-zero mean in the $y$ direction, indicating a transverse swimming velocity to the flow plane, being compatible with the bacterial bulk rheotaxis [Fig.\ref{pxpypz}(a)]. When $\Pe$ increases, as the orientational distribution is attracted towards the stable fixed point, the transverse components increase on average. 

The gyrotactic torques break the symmetry in the orientational distribution in $x$ and $z$ directions, yielding the non-zero mean values in $p_x$ and $p_z$ [Fig.\ref{pxpypz}(b)]. The chirality and gyrotactic effects are almost independent in this parameter set as shown in Fig.\ref{pxpypz}(c).

\section{Dispersion behaviours}
\label{sec:disp}

\subsection{Diffusion tensors}

\begin{figure}[t!]
\centering
\begin{overpic}[width=6cm]{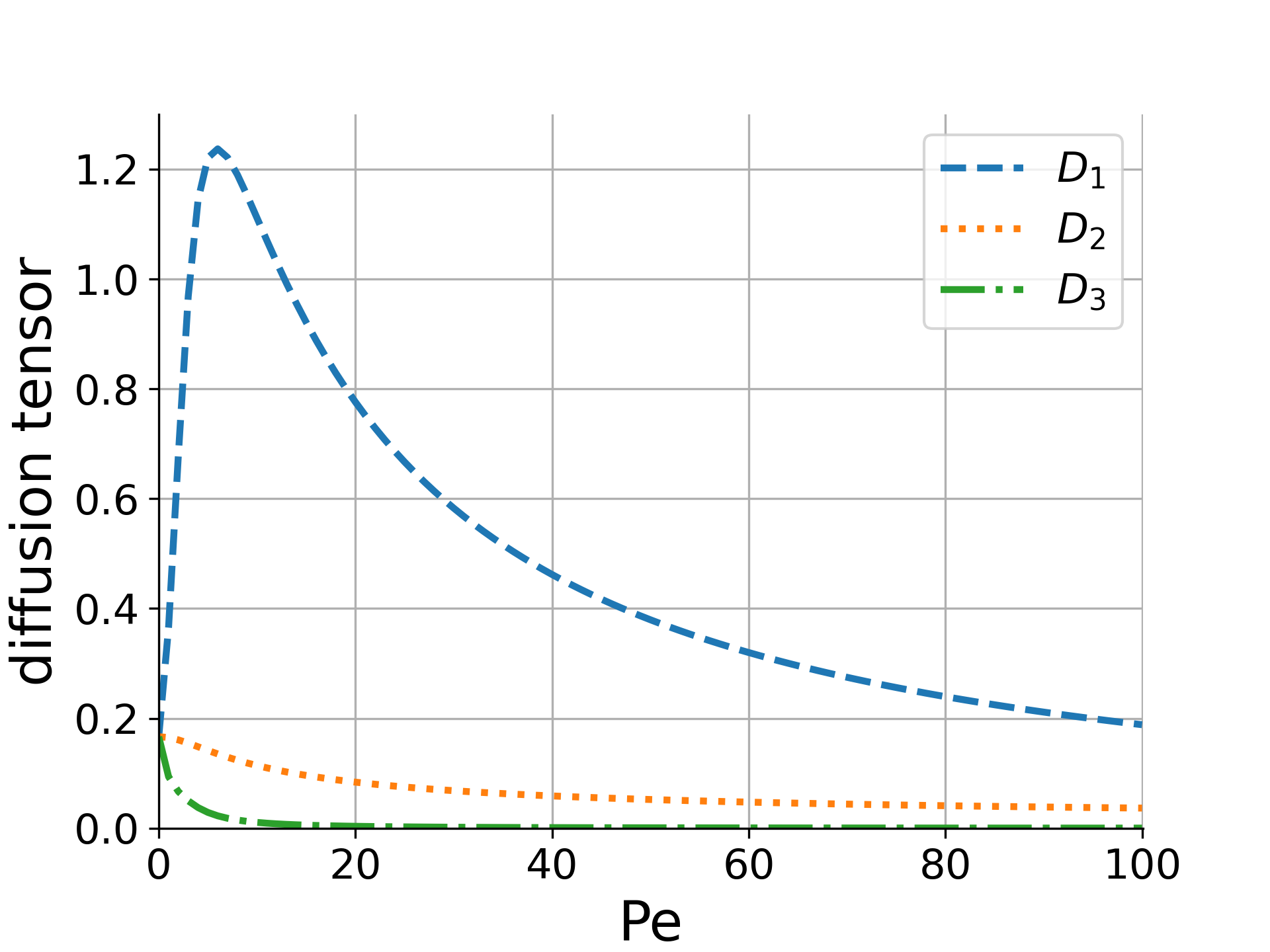}
\put(20,68){(a) ~$g=0,b=0.95,c=0$}
\end{overpic}
\begin{overpic}[width=6cm]{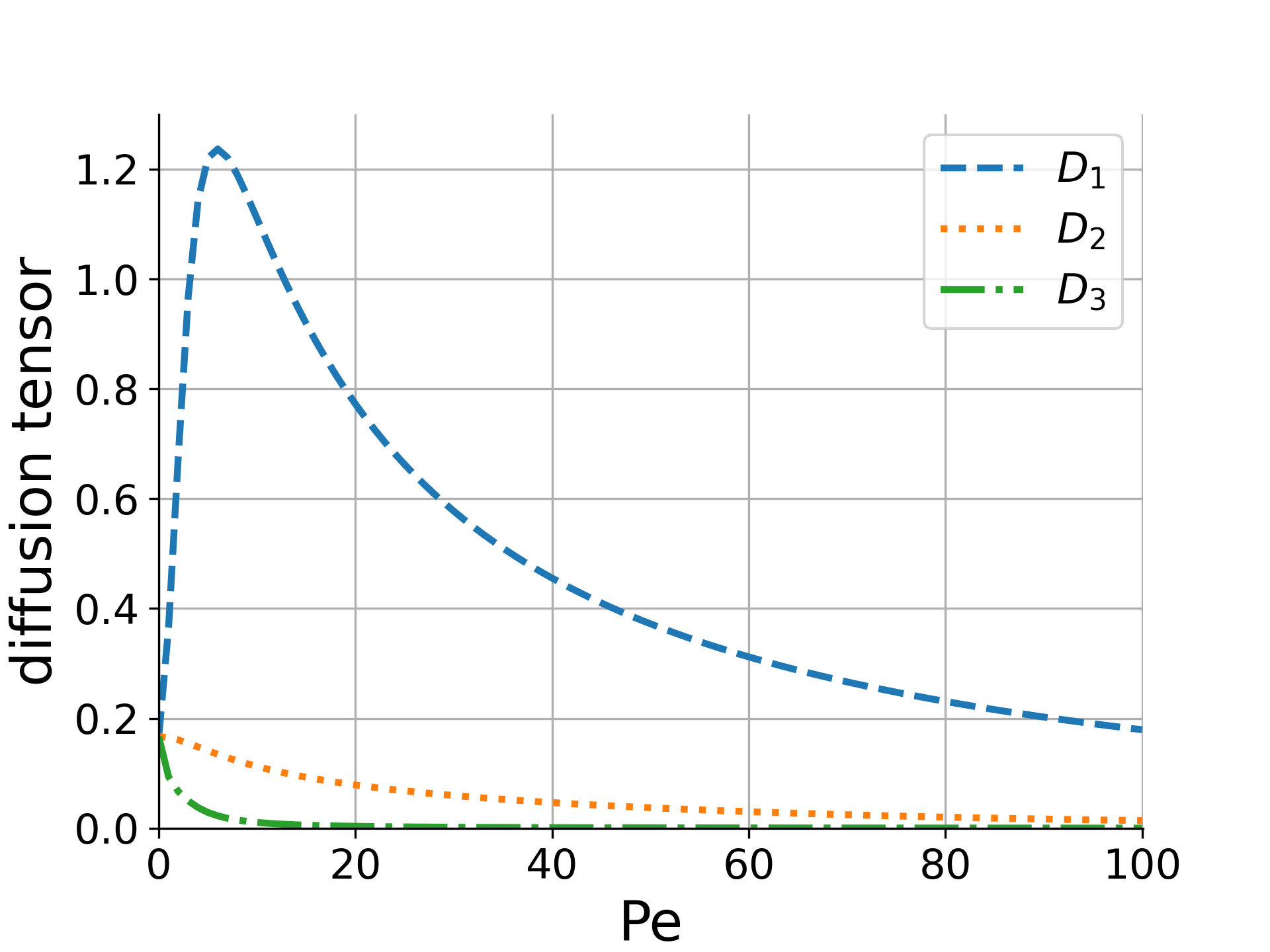}
\put(20,68){(b) ~$g=0,b=0.95,c=0.1$}
\end{overpic}
    \caption{Eigenvalues of the diffusion tensor for (a) non-chiral and (b) chiral particles with different $\Pe$ in $0\leq\text{Pe}\leq 100$. The diffusion in the $D_2$ axis is strongly suppressed due to the chirality effects at the high $\Pe$ region.}
    \label{D/}
\end{figure}

We now proceed to compute Eq.\eqref{nb} to obtain $\bm{b}$ and then the diffusion tensor from Eq.\eqref{diffusion}. In Fig.\ref{D/}, we show numerical plots of the eigenvalues of the diffusion tensors for non-chiral and chiral particles with different $\Pe$. With the correction term, the diffusion tensor is kept positive-definite.

In both panels of the figure, the eigenvalue $D_1$ has the largest value and $D_3$ has the smallest value. The corresponding eigenvectors are almost along the $x$ and $z$ axes, respectively. This result is in agreement with the classical Taylor dispersion that the distribution is strongly distorted towards the flow direction and we found that the chirality does not alter the dispersion dynamics in the flow plane.

In contrast, the eigenvalue $D_2$, whose eigenvector is directed to the $y$ axis, is strongly suppressed due to the chirality effect. With the parameter values of the figure, the suppression is more enhanced as $\Pe$ increases. For example, at $\Pe=100$, the value of $D_2$ is decreased to 38\% % {\bl [38 \% (37.9 \%)]}
for the parameter set used in Fig. \ref{D/}. This may be physically interpreted as the focus of the distribution function in the $\theta-\phi$ plane by the attracting fixed point for a large $\Pe$. 

The gyrotactic torque, however, does not qualitatively alter the eigenvalues. Indeed, when $g=0.03$, the plots of the eigenvalues are almost identical and the difference cannot be visible (figure not shown), which is distinct from the apparent gyrotactic effects in the averaged orientation of Fig. \ref{P0c}. This difference may be understood by the fact that the gyrotactic torque breaks the symmetry of the dynamical system but retains the stability of the fixed points.

\subsection{Dispersion in physical space}

%\begin{itemize}
%    \item Re-define the particle distributions by the obtained Gaussian distribution
%    \item Figure. Separation of active chiral particles
%    \item Figure. Separation of passive chiral particles
%\end{itemize}

The GTD theory assumes a Gaussian distribution for the population dynamics of microswimmers. With numerically obtained macroscopic model parameters $\bar{\bm{p}}$ and $\bar{\mathsf{D}}$, which are both constants independent of position and time, we may estimate dispersion behaviours in the physical space. %By assuming $\beta=\gamma=\delta=0$ as in the previous sections, we have the averaged velocity in the laboratory coordinates as $V_{s}\bar{\bm{p}}+Gz\bm{e}_{x}$. 
We set the initial position of the swimmers at the origin without loss of generality to write the mean position of the population as
$\bm{\mu}=(
        (1/2)GV_{s}A_{1}^{0}t^2+V_{s}A_{1}^{1}t,
        V_{s}B_{1}^{1}t,
        V_{s}A_{1}^{0}t
    )^{\textrm{T}}$
The variance matrix is then given by $2\mathsf{D}t=2(V_{s}^2/d_{r})\bar{\mathsf{D}}t$. The distribution function of the microswimmers that initially possesses a point-like distribution, given by the Dirac delta function $\delta(\bm{x})$ at time $t=0$, is then written as
\begin{equation}
    \bar{P}(\bm{x},t)=\frac{1}{\sqrt{(2\pi)^3 \textrm{det}(2\mathsf{D}t)}}\text{exp}\left[-\frac{1}{2}(\bm{x}-\bm{\mu})^{\text{T}}(2\mathsf{D}t)^{-1}(\bm{x}-\bm{\mu})\right]
    ~\textrm{for}~~t\geq0.
\end{equation}
%Let the eigenvalues of $\mathsf{D}$ be $D_{1},D_{2},D_{3}$ and corresponding normalised eigenvectors ve $x_{1},x_{2},x_{3}$, and we have
%\begin{equation}
%    \bar{P}(\bm{x},t)=\frac{1}{4\pi\sqrt{D_{1}D_{2}D_{3}\pi t}}\text{exp}\left[-\frac{1}{4t}\left(\frac{x_{1}^2}{D_{1}}+\frac{x_{2}^2}{D_{2}}+\frac{x_{3}^2}{D_{3}}\right)\right].
%\end{equation}

\begin{figure}[t!]
\centering
\begin{overpic}[width=4.4cm]{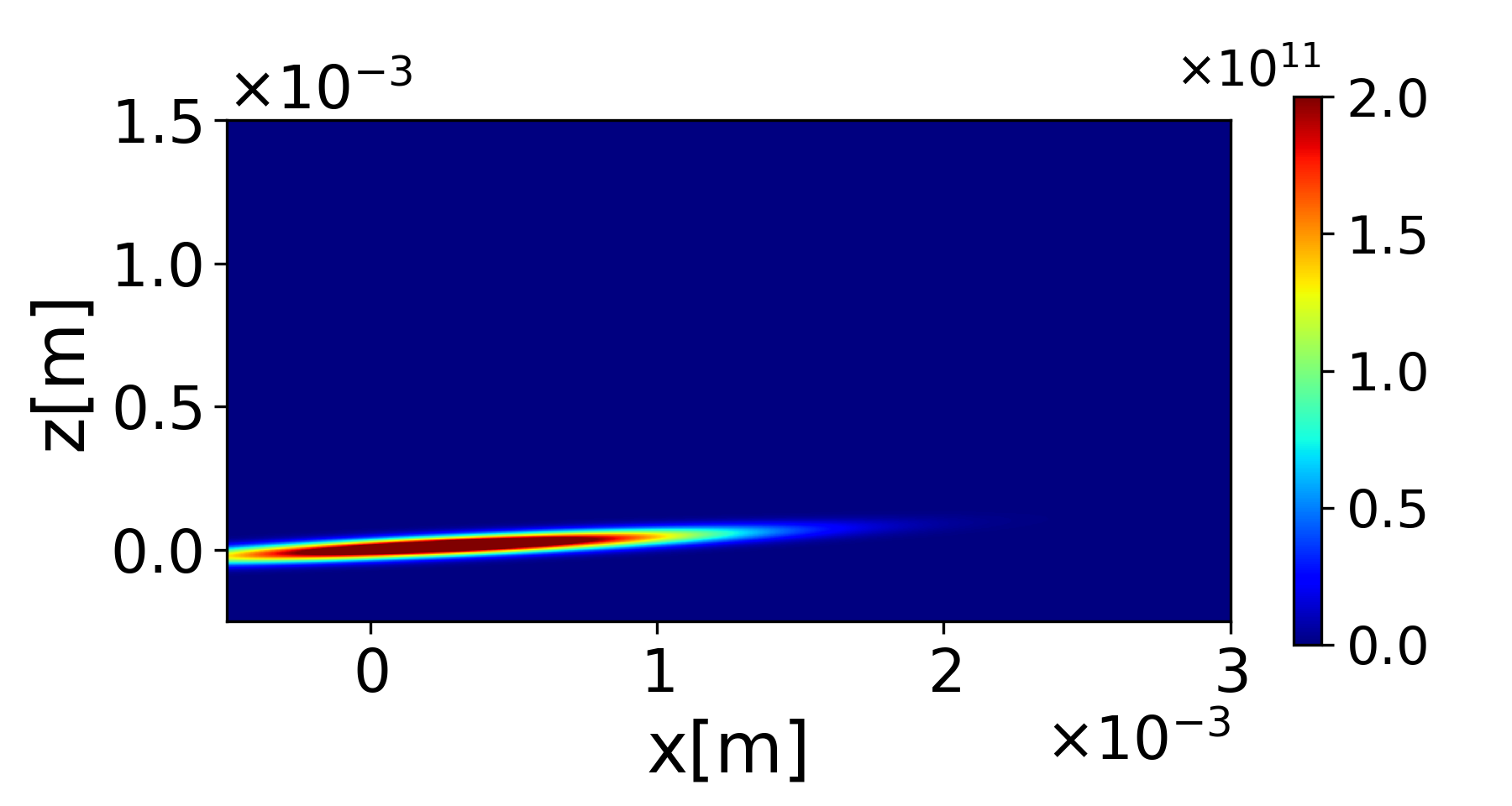}
\put(10,51){(a) ~$G=5\,\text{s}^{-1},t=5\,\text{s}$}
\end{overpic}
\begin{overpic}[width=4.4cm]{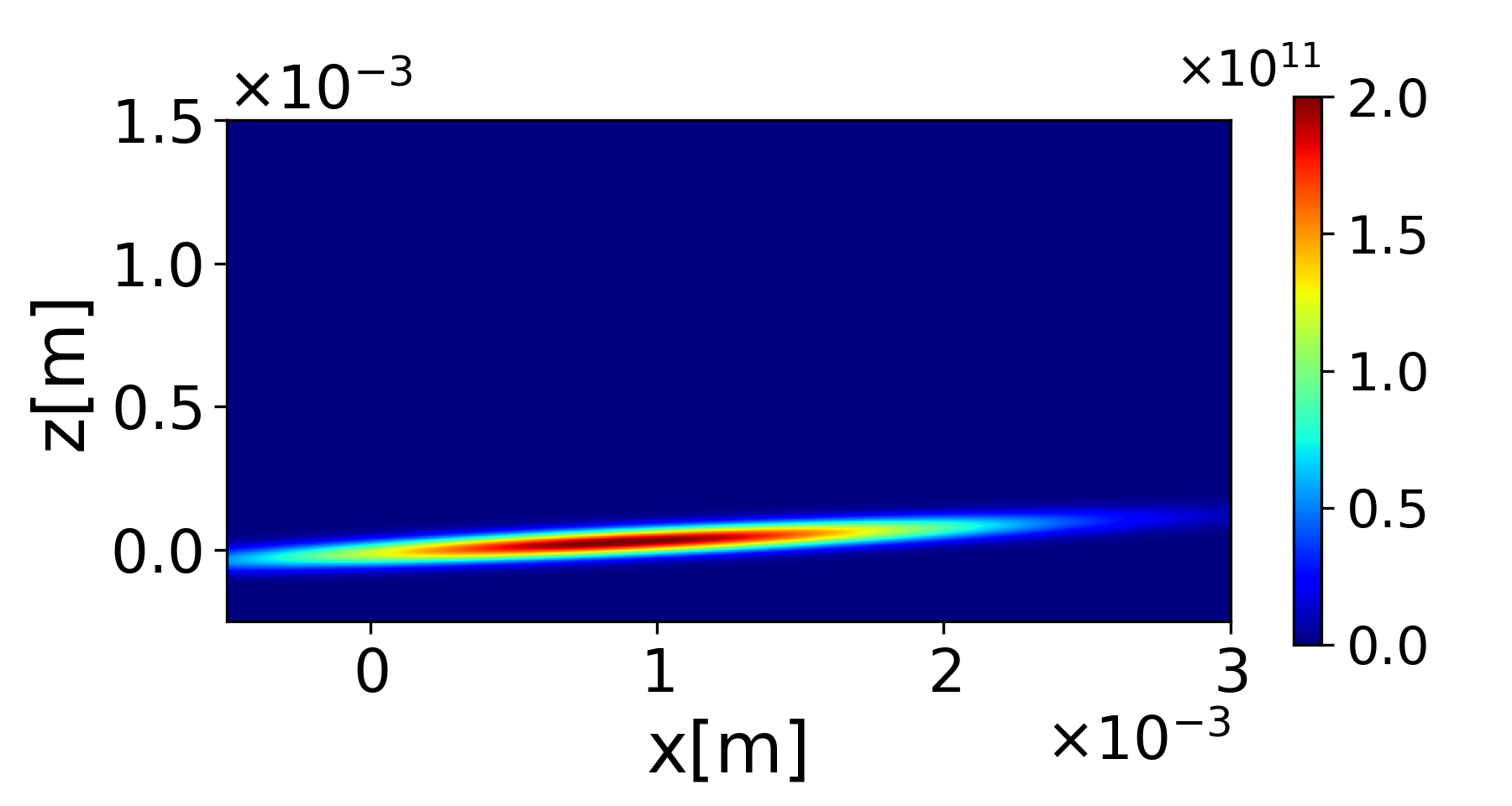}
\put(10,51){(b) ~$G=5\,\text{s}^{-1},t=10\,\text{s}$}
\end{overpic}
\begin{overpic}[width=4.4cm]{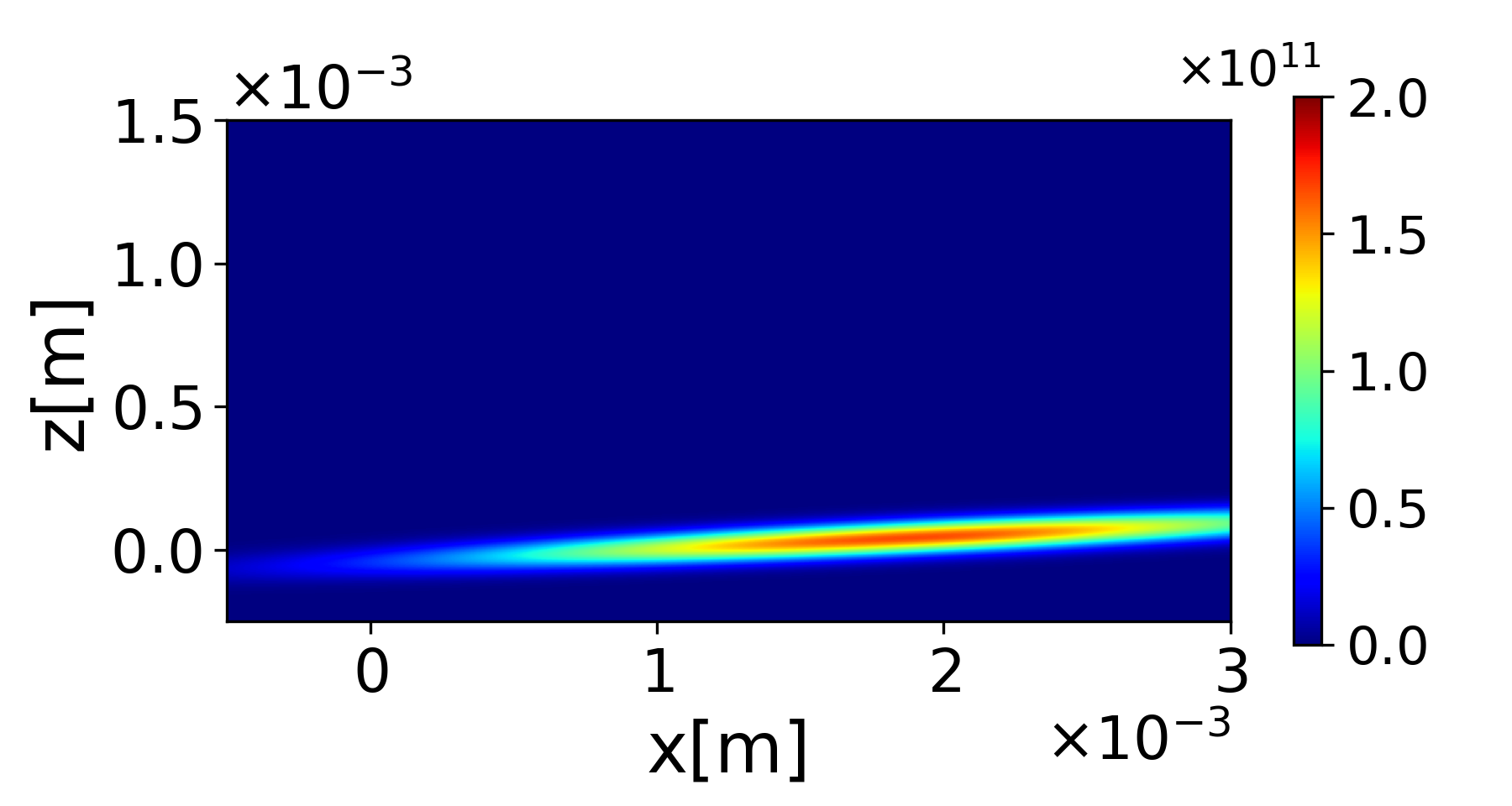}
\put(10,51){(c) ~$G=5\,\text{s}^{-1},t=15\,\text{s}$}
\end{overpic}\\
\begin{overpic}[width=4.4cm]{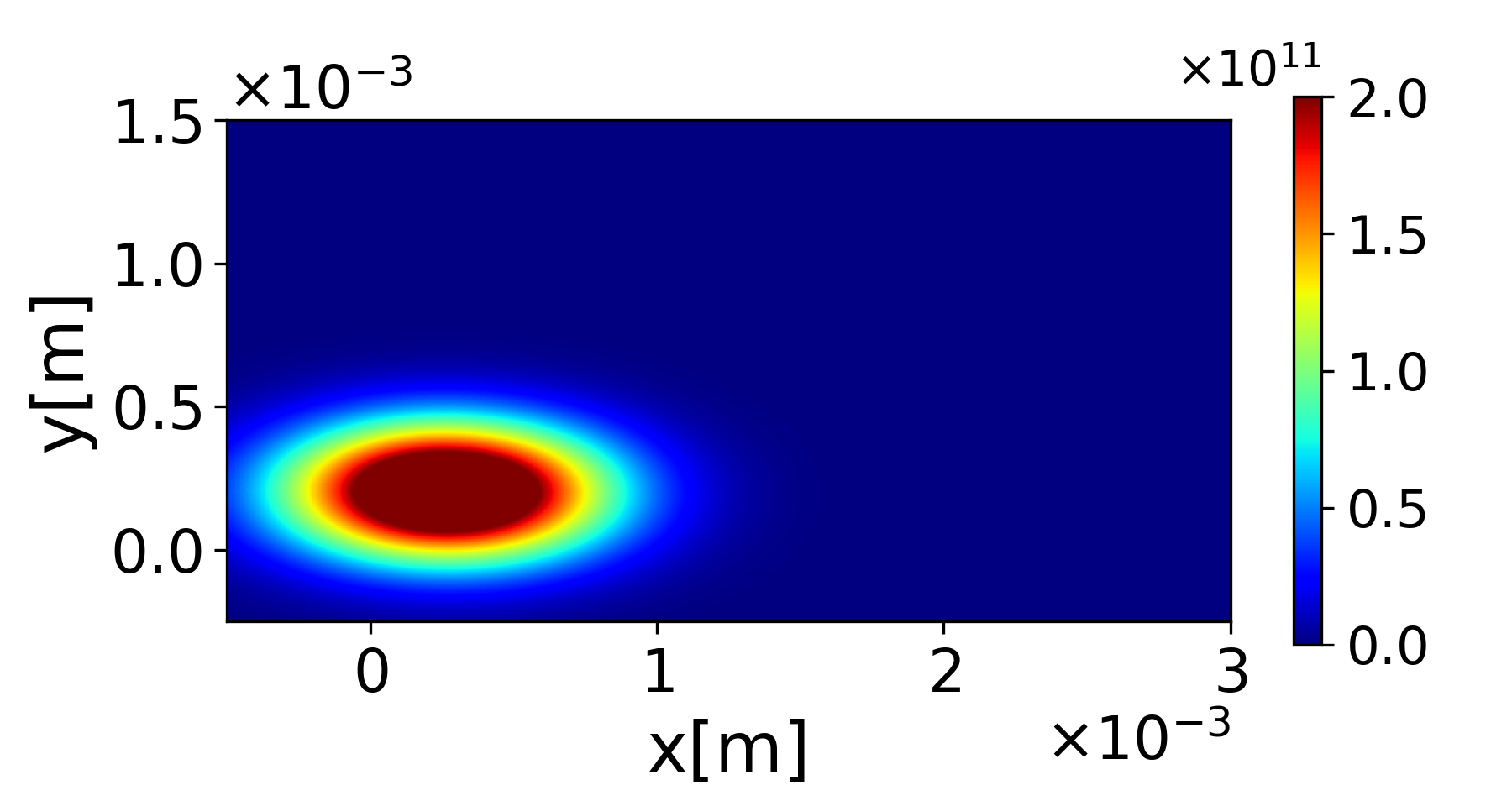}
\end{overpic}
\begin{overpic}[width=4.4cm]{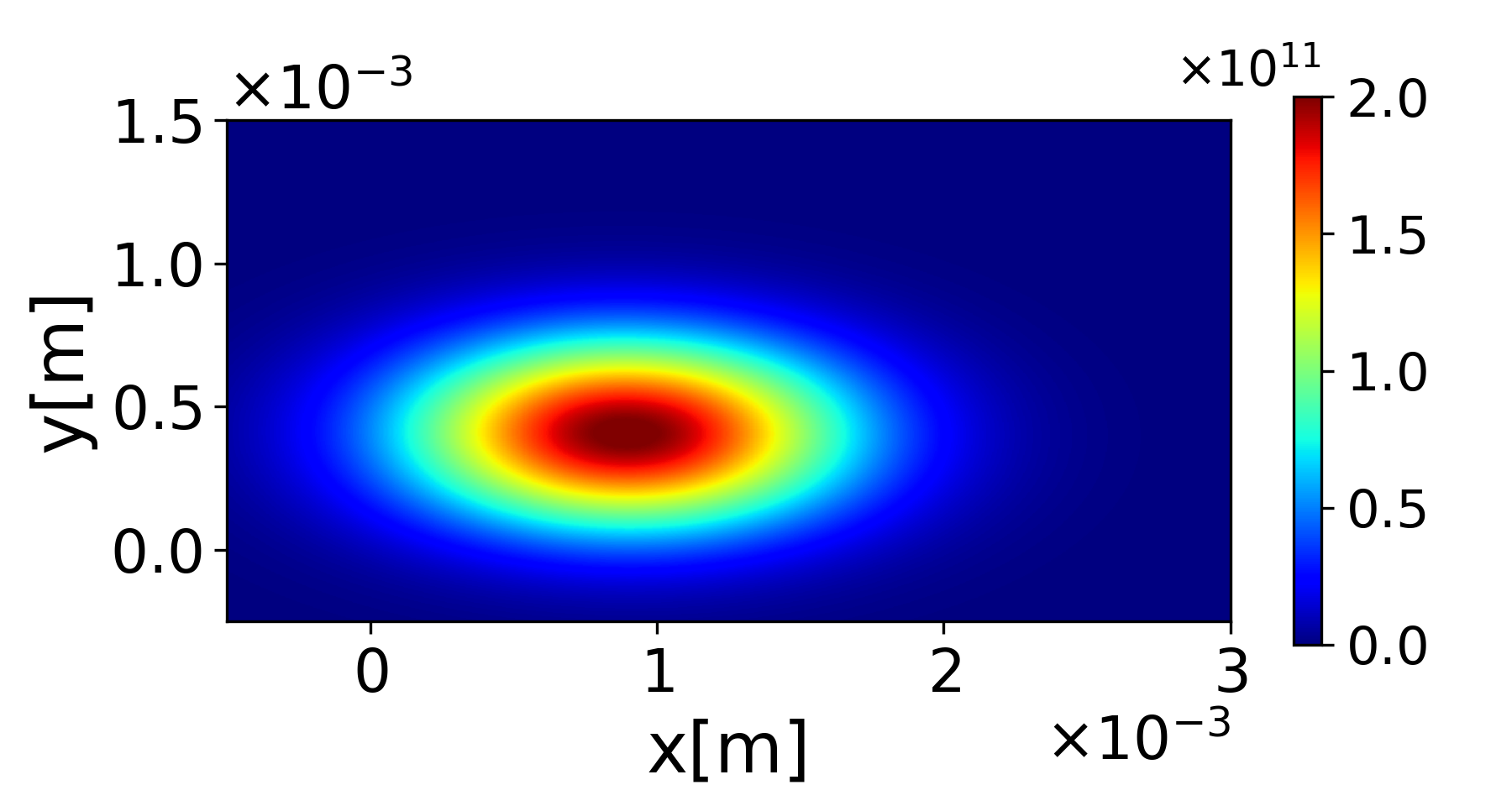}
\end{overpic}
\begin{overpic}[width=4.4cm]{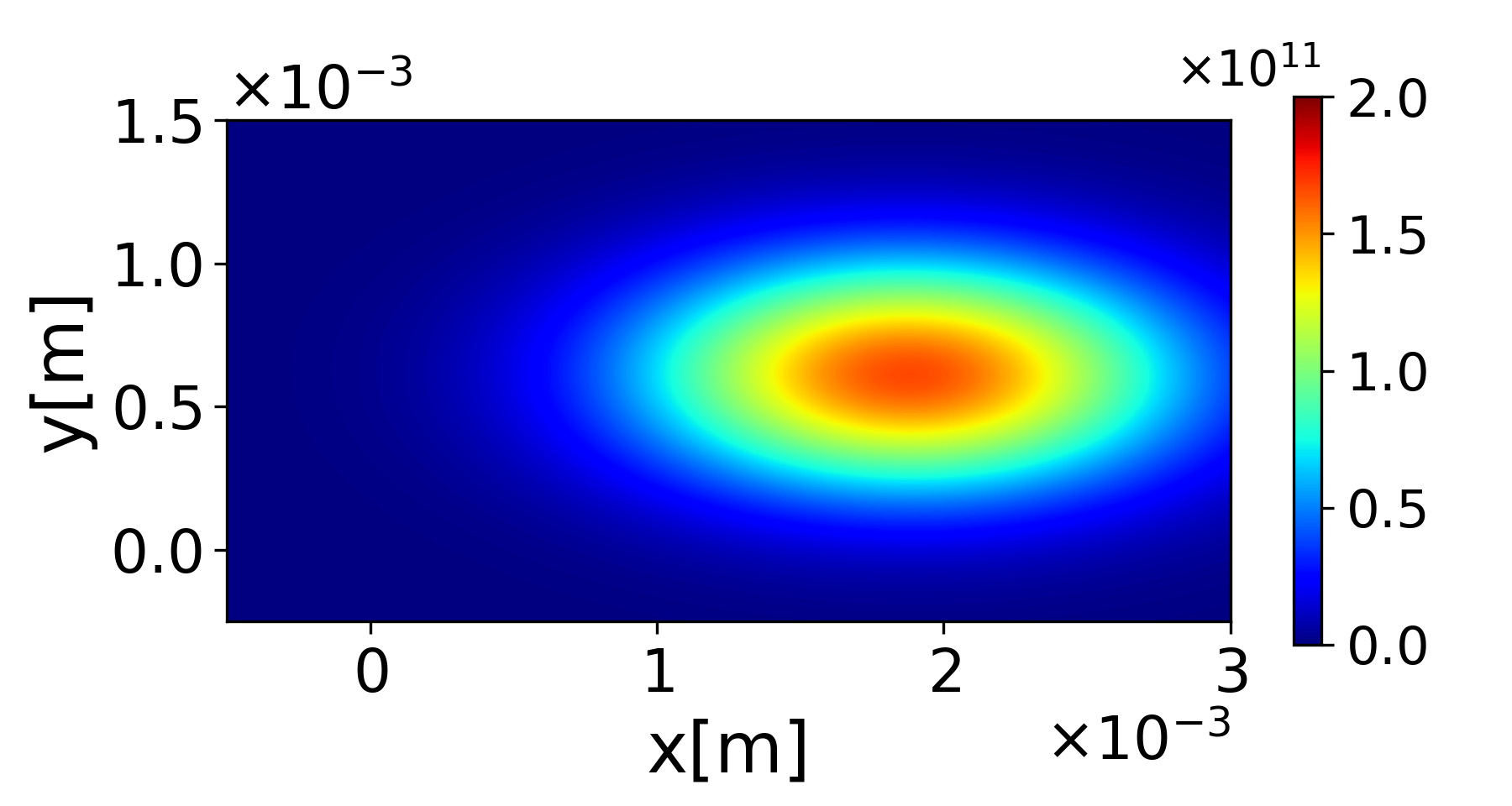}
\end{overpic}
    \caption{Snapshots of the distribution function for chiral microswimmers in simple shear at a  different time (a) $t=5$ sec, (b) $t=10$ sec, and (c) $t=15$ sec, in the in-flow plane ($x-z$ cross-section, top) and the $x-y$ cross-section (bottom). The parameter values are  $g=0.03,b=0.95,c=0.1, \text{Pe}=100$ and $G=5\,\text{s}^{-1}$.}
    \label{gaus}
  \end{figure}

In Fig.\ref{gaus}, we show snapshots of the distribution of chiral microswimmers with $g=0.03,b=0.95,c=0.3$ at $\text{Pe}=100$ with different time $t$. The cross-section in the $x$--$y$  and $x$--$z$ planes were chosen to contain the mean of the distribution, i.e., $z=V_{s}A_{1}^{0}t$ and $y=V_{s}B_{1}^{1}t$, respectively.
To illustrate the physical realisation of the population dynamics in a laboratory experiment, we set $V_{s}=10^{-4}\,\text{m/s}$ and $G=5\,\text{s}^{-1}$ with $t=5,10,15\,\text{s}$ and visualise the distribution function by a colour map with the physical units. %このとき，回転拡散係数$d_{r}$は$G$の値に対応して，$d_{r}=G/\text{Pe}=0.1,0.5\,\text{s}^{-1}$となる．

In the top panel of Fig.\ref{gaus}, we display the population of the microswimmers in the $x-z$ cross-section, on which the simple shear is applied. The swimmers are flown down by the shear and the distribution exhibits a large dispersion along the flow direction. The bottom panel of Fig.\ref{gaus} shows the same snapshots but in the $x-y$ cross-section. By the transverse velocity of the chiral microswimmers, the population moves towards the $y$ axis.  The suppression of the diffusion in the $y$ axis also contributes to focused rheotactic transport of the collection of the microswimmers.  If the swimmer possesses the opposite chirality $(c<0)$, the swimming direction becomes reversed. Hence, the mixture of chiral swimmers with the opposite chirality could be clearly separated in this simple shear.

%{\rd [Finally, we have explored the transport of passive chiral particles by using the shear-induced drift velocity. However, the effects were negligibly small as expected. Hence, we omit these results from the main text, though it may be worth mentioning somewhere in the text.] }

%{\rd [Check: Is $\beta\neq 0$ really negligibly small for the results of $\bar{\bm{U}}$ and $D_{ij}$? It is enough to numerically check for a small $\Pe$ and for a value of $\beta$ in E.coli bacteria. Checked.]}

\begin{comment}
\begin{figure}[t!]
\centering
\begin{overpic}[width=4.4cm]{gauss/Pe=10_G=5/xz_g=0.03_b=0.95_c=0.1_Pe=10_G=5_t=5.png}
\put(10,45){(a) ~$G=5\,\text{s}^{-1},t=5\,\text{s}$}
\end{overpic}
\begin{overpic}[width=4.4cm]{gauss/Pe=10_G=5/xz_g=0.03_b=0.95_c=0.1_Pe=10_G=5_t=10.png}
\put(10,45){(b) ~$G=5\,\text{s}^{-1},t=10\,\text{s}$}
\end{overpic}
\begin{overpic}[width=4.4cm]{gauss/Pe=10_G=5/xz_g=0.03_b=0.95_c=0.1_Pe=10_G=5_t=15.png}
\put(10,45){(c) ~$G=5\,\text{s}^{-1},t=15\,\text{s}$}
\end{overpic}
\begin{overpic}[width=4.4cm]{gauss/Pe=10_G=5/xy_g=0.03_b=0.95_c=0.1_Pe=10_G=5_t=5.png}
\end{overpic}
\begin{overpic}[width=4.4cm]{gauss/Pe=10_G=5/xy_g=0.03_b=0.95_c=0.1_Pe=10_G=5_t=10.png}
\end{overpic}
\begin{overpic}[width=4.4cm]{gauss/Pe=10_G=5/xy_g=0.03_b=0.95_c=0.1_Pe=10_G=5_t=15.png}
\end{overpic}
    \caption{$g=0.03,b=0.95,{\bl c=0.1, \text{Pe}=10} $ $G=5\,\text{s}^{-1}$. {\rd [Similar figures for a non-motile chiral particles]}}
    \label{gaus}
  \end{figure}
\end{comment}

\section{Conclusion}
\label{sec:conc}

In this study, we have numerically investigated the dispersion phenomena of chiral microswimmers in a vertical shear flow based on the generalised Taylor dispersion (GTD) theory, focusing on the shear-induced torque effects due to the chirality of the particles. With a spherical harmonic expansion, we were able to calculate macroscopic parameters of population-level swimmer transportation including the averaged swimming velocity and effective diffusion tensor. In particular, at a high $\Pe$ regime, higher-order spherical harmonics are necessary to tackle non-spherical, chiral particles, which individually obey an extended Jeffery equation with particle chirality.

With the chiral effects of the swimmer shape, the shear-induced torque aligns the swimmer towards the background vorticity, which leads to perpendicular migration to the shear plane as well studied for bacterial bulk rheotaxis.

This biased locomotion is well understood by the individual deterministic dynamical systems as an attracting fixed point in the orientational phase space. When the Pecl\'et number $\Pe$ is small ($\Pe \lesssim O(10^1))$, however, the chiral torques are masked by individual rotational diffusion and the orientational distribution is almost symmetric to the flow plane. As the Pecl\'et number increased ($\Pe=O(10^2)$), however, the distribution becomes biased to exhibit bimodal peaks around the attracting fixed point until it accumulates to a single peak at a large $\Pe$. 
The qualitative transitions are in good agreement with experimental and particle-based numerical simulations of E.coli bacteria \cite{Jing2020}.

With correction terms from the background shear, we were able to guarantee the positive-definiteness of the diffusion tensor and found that the diffusion coefficient $D_2$ in the direction perpendicular to the flow plane is remarkably reduced by the chiral effect, which suggests the flow-mediated cell separation depending on the sign of the particle chirality.
Indeed, the GTD theory also enables us to reconstruct the population-level distribution and our numerical simulations with typical parameter sets demonstrated that microswimmers can be separated out depending on their chirality. These results emphasise the usefulness of the GTD theory to predict and control the population-level dispersion swimmer behaviours and to design a tailor-made microdevice for cell sorting.

To apply our computational tools to the actual swimmers and particles, we need to estimate the chirality-induced drift and torque, which appear as the shape parameters in individual deterministic dynamical systems. The relation between the actual physical shape and hydrodynamic shape parameters is not trivial and thus demands further study \cite{Moreau2022, Dalwadi2024a, Dalwadi2024b}. Also, we have neglected the chirality-induced drift terms, since these are significantly small to the particle's swimming velocity. For a passive particle, however, the chirality-induced drift terms have to be included and indeed, this drift is considered to separate particles with the opposite chirality \cite{Makino2005, Eichhorn2010, Ro2016, Zottl2023}.

Nonetheless, the GTD theory is theoretically formulated in a linear, non-extensional flow and other complex flows such as a Poisseulle flow need special care to be applied.  Bees \& Croze\cite{BC2010,BC2012} studied its application to a pipe flow, and found that the GTD theory could be validated when the swimming velocity and shear variance are sufficiently small. Bearon, Hazel \& Thorn\cite{Bearon2011} examined an extension of the GTD theory in an extensional flow by comparing it with stochastic simulation, although the theory can accurately predict the collective behaviours only in one-dimensional flow. More recently, Fung, Bearon \& Hwang \cite{Fung2020,Fung2022} proposed local swimming velocity and local diffusion tensor, instead of modelling the dispersion by a global macroscopic parameter. The dispersion of chiral microswimmers in such flows is practically important in designing the microdevices and understanding the large-scale dynamics such as biocovection, which calls for further studies.

In summary, in this study, by using the GTD theory we have proposed theoretical and numerical tools to quantitatively analyse the dispersion phenomena of chiral microswimming such as bacteria cells. Our methods are useful for predicting and controlling  population-level dispersion and cell sorting.

\section*{Acknowledgments}
K.I. acknowledges the Japan Society for the Promotion of Science (JSPS) KAKENHI for Transformative Research Areas A (Grant No. 21H05309) and the Japan Science and Technology Agency (JST), FOREST (Grant No. JPMJFR212N). 

\appendix
\section{Appendix: Details of calculations}
\label{app}

In this appendix section, we provide some details of the numerical computations. To start, we first introduce the following expressions for later calculations,
\begin{alignat}{2}
    &F_{n}^{m}(\theta,\phi)&\equiv& [A_{n}^{m}\cos m\phi+B_{n}^{m}\sin m\phi] P_{n}^{m}(\cos\theta)\label{Fnm},\\
    &F_{nj}^{m}(\theta,\phi)&\equiv& [\alpha_{nj}^{m}\cos m\phi+\beta_{nj}^{m}\sin m\phi]P_{n}^{m}(\cos\theta),\\
    &R_{n}^{m}(\phi)&\equiv& A_{n}^{m}\cos m\phi+B_{n}^{m}\sin m\phi,\\
    &R_{nj}^{m}(\phi)&\equiv& \alpha_{nj}^{m}\cos m\phi+\beta_{nj}^{m}\sin m\phi,\\
    &Q_{n}^{m}(\theta,\phi)&\equiv& \cos m\phi P_{n}^{m}(\cos\theta),\\
    &S_{n}^{m}(\theta,\phi)&\equiv& \sin m\phi P_{n}^{m}(\cos\theta) \label{Snm}.
\end{alignat}
With these, the orientational vector $\bm{p}$ is simply given by
\begin{equation}
    \bm{p}=\begin{pmatrix}
    \sin\theta\cos\phi \\ \sin\theta\sin\phi \\ \cos\theta
    \end{pmatrix}=
    \begin{pmatrix}
    Q_{1}^{1} \\ S_{1}^{1} \\ Q_{1}^{0}
    \end{pmatrix}.
\end{equation}
Also, it is useful to recall the orthonormal relations of the associated Legendre polynomials
\begin{align}
    &\int_{\mathbb{S}^2}Q_{n}^{m}Q_{n^{\prime}}^{m^{\prime}}d^2\bm{p}=\delta_{mm^{\prime}}\delta_{nn^{\prime}}\frac{2\pi}{2n+1}\frac{(n+m)!}{(n-m)!}\quad\text{for} \quad n,n^{\prime}\geq m,m^{\prime}=1,2,\ldots\quad,\\
    &\int_{\mathbb{S}^2}Q_{n}^{0}Q_{n^{\prime}}^{m^{\prime}}d^2\bm{p}=\delta_{0m^{\prime}}\delta_{nn^{\prime}}\frac{4\pi}{2n+1}\quad\text{for} \quad n,n^{\prime}\geq m^{\prime}=1,2,\ldots,
\end{align}
where $\delta_{nm}$ indicates the Kronecker delta.

By substituting Eqs.\eqref{Fnm}-\eqref{Snm} into the expression of the operator $\mathcal{L}$ \eqref{L},  
The equation for $P_0^\infty$ in Eq. \eqref{beq} may be rewritten as
\begin{align}
\begin{split}
    \sum_{n=0}^{\infty}\sum_{m=0}^{n}\biggl(&
    n(n+1)F_{n}^{m}+\frac{1}{2}\text{Pe}\bigg[\{g\sin\theta-(1+b\cos 2\theta)\cos\phi+c\cos\theta\sin\phi\}\sin\theta R_{n}^{m}P_{n}^{m\prime}\\
    &-\frac{(1+b)\cos\theta\sin\phi+c\cos 2\theta\cos\phi}{\sin\theta} R_{n}^{m\prime}P_{n}^{m}
    -(2g\cos\theta+6b\sin\theta\cos\theta\cos\phi)F_{n}^{m}\biggl]\bigg)=0
\end{split}
\label{P0infinity}.
\end{align}
The normalisation constraint is rephrased into
\begin{align}
    A_{0}^{0}=\frac{1}{4\pi}
\end{align}

To determine the coefficients of the spherical harmonics, we need further mathematical manipulations. To do so, let us first summarise useful formulae on the associated Legendre polynomials, $P_{n}^{m}(\cos\theta)$, 
\begin{align}
    \cos\theta P_{n}^{m}&\equiv\frac{n+m}{2n+1}P_{n-1}^{m}+\frac{n-m+1}{2n+1}P_{n+1}^{m},\\
    \sin\theta P_{n}^{m}&\equiv\frac{1}{2n+1}(P_{n+1}^{m+1}-P_{n-1}^{m+1}),\\
    \sin\theta P_{n}^{m}&\equiv\frac{1}{2n+1}\left[(n+m)(n+m-1)P_{n-1}^{m-1}-(n-m+1)(n-m+2)P_{n+1}^{m-1}\right],\\
    \sin\theta P_{n}^{m\prime}&\equiv P_{n}^{m+1}-\frac{m\cos\theta}{\sin\theta}P_{n}^{m},\\
    \sin\theta P_{n}^{m\prime}&\equiv -(n-m+1)(n+m)P_{n}^{m-1}+\frac{m\cos\theta}{\sin\theta}P_{n}^{m},\\
    \sin^2\theta P_{n}^{m\prime}&\equiv\frac{1}{2n+1}\left[(n+m)(n+1)P_{n-1}^{m}-n(n-m+1)P_{n+1}^{m}\right].
\end{align}
We now use these identities to represent  Eq.\eqref{P0infinity} in the basis $Q_n^m$ and $P_n^m$. After some cumbersome calculations, we may rewrite the left-hand side of Eq.\eqref{P0infinity} in the form,
\begin{align}
    %\sum_{n=0}^{\infty}\sum_{m=0}^{n}\biggl(&
    %n(n+1)F_{n}^{m}+\frac{1}{2}\text{Pe}\bigg[\{g\sin\theta-(1+b\cos 2\theta)\cos\phi+c\cos\theta\sin\phi\}\sin\theta R_{n}^{m}P_{n}^{m\prime}\\
    %&-\frac{(1+b)\cos\theta\sin\phi+c\cos 2\theta\cos\phi}{\sin\theta} R_{n}^{m\prime}P_{n}^{m}-(2g\cos\theta+6b\sin\theta\cos\theta\cos\phi)F_{n}^{m}\biggl]\bigg)\\
    %=&
    \sum_{n=0}^{\infty}\sum_{m=0}^{n}\left( L_n^m Q_n^m+M_n^m S_n^m\right)=0,
\end{align}
where the coefficients, $L_n^m$ and $M_n^m$ are explicitly given by
\begin{align}     
    L_n^m=
    &-\frac{b(n-m)(n+1)}{(2n-3)(2n-1)}A_{n-2}^{m-1}+\frac{b(n-m-2)(n-m-1)(n-m)(n+1)}{(2n-3)(2n-1)}A_{n-2}^{m+1}  \nonumber  \\
    &-\frac{g(n+1)(n-m)}{2n-1}A_{n-1}^{m}+\bigg[-\frac{1}{2}+b\bigg\{\frac{(n+m+1)(n-2)}{(2n-1)(2n+1)}+\frac{(n-m+2)(n+3)}{(2n+1)(2n+3)}-\frac{1}{2}\bigg\}\bigg]A_{n}^{m-1}\nonumber \\
    &+n(n+1)A_{n}^{m}+\bigg[\frac{(n-m)(n+m+1)}{2}+b\bigg\{-\frac{(n+m+1)(n-m-1)(n-m)(n+2)}{(2n-1)(2n+1)}\nonumber \\
    &-\frac{(n-m)(n+m+2)(n+m+1)(n+3)}{(2n+1)(2n+3)}+\frac{(n-m)(n+m+1)}{2}\bigg\}\bigg]A_{n}^{m+1}\nonumber \\
    &+\frac{gn(n+m+1)}{2n+3}A_{n+1}^{m}-\frac{bn(n+m+1)}{(2n+3)(2n+5)}A_{n+2}^{m-1}+\frac{b(n+m+3)(n+m+2)(n+m+1)n}{(2n+3)(2n+5)}A_{n+2}^{m+1}\nonumber \\
    &-\frac{c(n-2m+1)}{2(2n-1)}B_{n-1}^{m-1}-\frac{c(n-m)(n-m-1)(n+2m+1)}{2(2n-1)}B_{n-1}^{m+1}\nonumber \\
    &-\frac{c(n+2m)}{2(2n+3)}B_{n+1}^{m-1}-\frac{c(n+m+2)(n+m+1)(n-2m)}{2(2n+3)}B_{n+1}^{m+1}
    \label{Q}
\end{align}
and 
\begin{align}
    M_n^m=
    &\frac{c(n-2m+1)}{2(2n-1)}A_{n-1}^{m-1}+\frac{c(n-m)(n-m-1)(n+2m+1)}{2(2n-1)}A_{n-1}^{m+1} \nonumber \\
    &+\frac{c(n+2m)}{2(2n+3)}A_{n+1}^{m-1}+\frac{c(n+m+2)(n+m+1)(n-2m)}{2(2n+3)}A_{n+1}^{m+1}  \nonumber \\
    &-\frac{b(n-m)(n+1)}{(2n-3)(2n-1)}B_{n-2}^{m-1}+\frac{b(n-m-2)(n-m-1)(n-m)(n+1)}{(2n-3)(2n-1)}B_{n-2}^{m+1}  \nonumber \\
    &-\frac{g(n+1)(n-m)}{2n-1}B_{n-1}^{m}+\bigg[-\frac{1}{2}+b\bigg\{\frac{(n+m+1)(n-2)}{(2n-1)(2n+1)}+\frac{(n-m+2)(n+3)}{(2n+1)(2n+3)}-\frac{1}{2}\bigg\}\bigg]B_{n}^{m-1}  \nonumber \\
    &+n(n+1)B_{n}^{m}+\bigg[\frac{(n-m)(n+m+1)}{2}+b\bigg\{-\frac{(n+m+1)(n-m-1)(n-m)(n+2)}{(2n-1)(2n+1)} \nonumber \\
    &-\frac{(n-m)(n+m+2)(n+m+1)(n+3)}{(2n+1)(2n+3)}+\frac{(n-m)(n+m+1)}{2}\bigg\}\bigg]B_{n}^{m+1} \nonumber \\
    &+\frac{gn(n+m+1)}{2n+3}B_{n+1}^{m}-\frac{bn(n+m+1)}{(2n+3)(2n+5)}B_{n+2}^{m-1} \nonumber \\
    &+\frac{b(n+m+3)(n+m+2)(n+m+1)n}{(2n+3)(2n+5)}B_{n+2}^{m+1}
\label{S}.
\end{align}
Each coefficient, $L_{n}^{m}$ and $M_{n}^{m}$, are identically zeros, which forms a linear problem to determine $A_n^m$ and $B_n^b$, by which we can obtain $P_0^\infty$.

Similarly, we may rewrite the equation for $\bm{b}$ in Eq. \eqref{beq}. Noting the $\mathcal{L}$ operator is in common in the three equations for $\bm{b}$, we introduce $H_1$, $H_2$ and $H_3$ as
\begin{align}
\begin{split}
    H_i=\sum_{n=0}^{\infty}\sum_{m=0}^{n}\biggl(&
    n(n+1)F_{ni}^{m}+\frac{1}{2}\text{Pe}\bigg[\{g\sin\theta-(1+b\cos 2\theta)\cos\phi+c\cos\theta\sin\phi\}\sin\theta R_{n1}^{m}P_{n}^{m\prime}\\
    &-\frac{(1+b)\cos\theta\sin\phi+c\cos 2\theta\cos\phi}{\sin\theta} R_{n1}^{m\prime}P_{n}^{m}
    -(2g\cos\theta+6b\sin\theta\cos\theta\cos\phi)F_{ni}^{m}\biggl]\bigg)
\end{split}
\label{Hi},
\end{align}
and arrive at the following equations:
\begin{eqnarray}
H_1&=&\sum_{n=0}^{\infty}\sum_{m=0}^{n}\left[\text{Pe}\alpha_{n3}^{m}Q_{n}^{m}+\text{Pe}\beta_{n3}^{m}S_{n}^{m}
    +\left(\sin\theta\cos\phi-\frac{4\pi}{3}A_{1}^{1}\right)(A_{n}^{m}Q_{n}^{m}+B_{n}^{m}S_{n}^{m})\right],
\label{b1}\\
H_2&=&\sum_{n=0}^{\infty}\sum_{m=0}^{n}\left(\sin\theta\sin\phi-\frac{4\pi}{3}B_{1}^{1}\right)(A_{n}^{m}Q_{n}^{m}+B_{n}^{m}S_{n}^{m}),
\label{b2} \\ 
H_3&=&\sum_{n=0}^{\infty}\sum_{m=0}^{n}\left(\cos\theta-\frac{4\pi}{3}A_{1}^{0}\right)(A_{n}^{m}Q_{n}^{m}+B_{n}^{m}S_{n}^{m}),
\label{b3}
\end{eqnarray}
together with the normalisation constraints,
\begin{align}
    \alpha_{0j}^{0}=0\quad(j=1,2,3).
\end{align}

We also rewrite these equations to be presented in the bases of $Q_n^m$ and $S_n^m$. The left-hand sides of Eqs. \eqref{b1}-\eqref{b3} are the same lengthy expressions of \eqref{Q}-\eqref{S}, after replacing $A_{n}^{m},B_{n}^{m}$ by $\alpha_{nj}^{m},\beta_{nj}^{m}$. The right-hand sides of Eqs. \eqref{b1}-\eqref{b3} may be calculated as
\begin{align}
    \begin{split}
        \sum_{n=0}^{\infty}\sum_{m=0}^{n}\bigg(&\text{Pe}\alpha_{n3}^{m}Q_{n}^{m}+\text{Pe}\beta_{n3}^{m}S_{n}^{m}
        +\left[\sin\theta\cos\phi-\frac{4\pi}{3}A_{1}^{1}\right](A_{n}^{m}Q_{n}^{m}+B_{n}^{m}S_{n}^{m})\bigg)\\
        =\sum_{n=0}^{\infty}\sum_{m=0}^{n}\bigg(
        &\frac{1}{2(2n-1)}A_{n-1}^{m-1}-\frac{(n-m-1)(n-m)}{2(2n-1)}A_{n-1}^{m+1}-\frac{4\pi}{3}A_{1}^{1}A_{n}^{m}\\
        &+\text{Pe}\alpha_{n3}^{m}-\frac{1}{2(2n+3)}A_{n+1}^{m-1}+\frac{(n+m+2)(n+m+1)}{2(2n+3)}A_{n+1}^{m+1}
        \bigg)\times Q_{n}^{m}\\
        +\sum_{n=0}^{\infty}\sum_{m=0}^{n}\bigg(
        &\frac{1}{2(2n-1)}B_{n-1}^{m-1}-\frac{(n-m-1)(n-m)}{2(2n-1)}B_{n-1}^{m+1}-\frac{4\pi}{3}A_{1}^{1}B_{n}^{m}\\
        &+\text{Pe}\beta_{n3}^{m}-\frac{1}{2(2n+3)}B_{n+1}^{m-1}+\frac{(n+m+2)(n+m+1)}{2(2n+3)}B_{n+1}^{m+1}
        \bigg)\times S_{n}^{m},
    \end{split}
    \label{b1S}\\
    \begin{split}
        \sum_{n=0}^{\infty}\sum_{m=0}^{n}\bigg[&\sin\theta\sin\phi-\frac{4\pi}{3}B_{1}^{1}\bigg](A_{n}^{m}Q_{n}^{m}+B_{n}^{m}S_{n}^{m})\\
        =\sum_{n=0}^{\infty}\sum_{m=0}^{n}\bigg(
        &-\frac{1}{2(2n-1)}B_{n-1}^{m-1}-\frac{(n-m-1)(n-m)}{2(2n-1)}B_{n-1}^{m+1}-\frac{4\pi}{3}B_{1}^{1}A_{n}^{m}\\
        &+\frac{1}{2(2n+3)}B_{n+1}^{m-1}+\frac{(n+m+2)(n+m+1)}{2(2n+3)}B_{n+1}^{m+1}
        \bigg)\times Q_{n}^{m}\\
        +\sum_{n=0}^{\infty}\sum_{m=0}^{n}\bigg(
        &\frac{1}{2(2n-1)}A_{n-1}^{m-1}+\frac{(n-m-1)(n-m)}{2(2n-1)}A_{n-1}^{m+1}-\frac{4\pi}{3}B_{1}^{1}B_{n}^{m}\\
        &-\frac{1}{2(2n+3)}A_{n+1}^{m-1}-\frac{(n+m+2)(n+m+1)}{2(2n+3)}A_{n+1}^{m+1}
        \bigg)\times S_{n}^{m},
    \end{split}
    \label{b2S}
\end{align}
and
\begin{align}
    \begin{split}
        &\sum_{n=0}^{\infty}\sum_{m=0}^{n}\left[\cos\theta-\frac{4\pi}{3}A_{1}^{0}\right](A_{n}^{m}Q_{n}^{m}+B_{n}^{m}S_{n}^{m})\\
        =&\sum_{n=0}^{\infty}\sum_{m=0}^{n}\bigg(
        \frac{n-m}{2n-1}A_{n-1}^{m}-\frac{4\pi}{3}A_{1}^{0}A_{n}^{m}+\frac{n+m+1}{2n+3}A_{n+1}^{m}
        \bigg)\times Q_{n}^{m}\\
        +&\sum_{n=0}^{\infty}\sum_{m=0}^{n}\bigg(
        \frac{n-m}{2n-1}B_{n-1}^{m}-\frac{4\pi}{3}A_{1}^{0}B_{n}^{m}+\frac{n+m+1}{2n+3}B_{n+1}^{m}
        \bigg)\times S_{n}^{m}
    \end{split}
    \label{b3S}
\end{align}
These identities among the coefficients for $Q_n^m$ and $S_n^m$ again form linear problems to determine the coefficients $\alpha_{nj}^m$ and $\beta_{nj}^m$, after substituting $A_n^m$ and $B_n^m$ which are obtained from \eqref{Q}-\eqref{S}.

%%%%%%%%%% Insert bibliography here %%%%%%%%%%%%%%

\end{document}